%% file: main.tex
\documentclass[acmlarge,authorversion]{acmart}

\AtBeginDocument{%
  \providecommand\BibTeX{{%
    \normalfont B\kern-0.5em{\scshape i\kern-0.25em b}\kern-0.8em\TeX}}}

\acmJournal{IMWUT}
\acmVolume{8}
\acmNumber{4}
\acmArticle{203}
\acmMonth{12}

\usepackage{caption}
\usepackage{subcaption}
\usepackage{_macros}
\usepackage{multirow}
\usepackage{color, colortbl}
\usepackage[title]{appendix}
\usepackage{url}
\usepackage{xcolor}
\usepackage{booktabs}
\usepackage{tabularx} %

\newcommand{\name}{EMooly}
\begin{document}
\setcopyright{licensedcagov}
\acmYear{2024} \acmVolume{8} \acmNumber{4} \acmArticle{203}
\acmMonth{12}\acmDOI{10.1145/3699738}

\title[\name]{\name: Supporting Autistic Children in Collaborative Social-Emotional Learning with Caregiver Participation through Interactive AI-infused and AR Activities}

\author{\href{https://orcid.org/0009-0008-7730-8552}{Yue Lyu}}
\email{yue.lyu@uwaterloo.ca}
\affiliation{
    \institution{University of Waterloo}
    \city{Waterloo}
    \state{ON}
    \country{Canada}
}

\author{\href{https://orcid.org/0009-0001-3513-6587}{Di Liu}}
\email{seucliudi@gmail.com}
\affiliation{
    \institution{Southern University of Science and Technology}
    \city{Shenzhen}
    \country{China}
}

\author{\href{https://orcid.org/0000-0002-7705-2031}{Pengcheng An}}
\email{anpc@sustech.edu.cn}
\affiliation{
    \institution{Southern University of Science and Technology}
    \city{Shenzhen}
    \country{China}
}

\author{\href{https://orcid.org/0000-0002-8037-6301}{Xin Tong}}
\email{xint@hkust-gz.edu.cn}
\affiliation{
    \institution{Hong Kong University of Science and Technology (Guangzhou)}
    \city{}
    \country{China}
}

\author{\href{https://orcid.org/0009-0001-0373-8241}{Huan Zhang}}
\email{h648zhang@uwaterloo.ca}
\affiliation{
    \institution{University of Waterloo}
    \city{Waterloo}
    \state{ON}
    \country{Canada}
}

\author{\href{https://orcid.org/0000-0002-9642-9666}{Keiko Katsuragawa}}
\email{kkatsuragawa@uwaterloo.ca}
\affiliation{
    \institution{National Research Council}
    \city{Waterloo}
    \state{ON}
    \country{Canada}
}
\affiliation{
    \institution{University of Waterloo}
    \city{Waterloo}
    \state{ON}
    \country{Canada}
}

\author{\href{https://orcid.org/0000-0001-5008-4319}{Jian Zhao}}
\email{jianzhao@uwaterloo.ca}
\affiliation{
    \institution{University of Waterloo}
    \city{Waterloo}
    \state{ON}
    \country{Canada}
}

\renewcommand{\shortauthors}{Lyu et al.}
\showrevisions{black}
\begin{abstract}
Children with autism spectrum disorder (ASD) have social-emotional deficits that lead to difficulties in recognizing emotions as well as understanding and responding to social interactions.
This study presents \name{}, a tablet game that actively involves caregivers and leverages augmented reality (AR) and generative AI (GenAI) to enhance social-emotional learning for autistic children. 
Through a year of collaborative effort with five domain experts, we developed \name{} that engages children through personalized social stories, interactive and fun activities, and enhanced caregiver participation, focusing on emotion understanding and facial expression recognition. 
Compared with a baseline, a controlled study with 24 autistic children and their caregivers showed \name{} significantly improved children's emotion recognition skills and its novel features were preferred and appreciated.
\name{} demonstrates the potential of AI and AR in enhancing social-emotional development for autistic children via prompt personalizing and engagement, and highlights the importance of caregiver involvement for optimal learning outcomes.
\end{abstract}




\begin{teaserfigure}
    \centering
    \includegraphics[width=1.\linewidth]{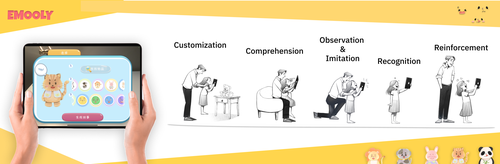}
    \vspace{-8mm}
    \caption{\name{} is an innovative tablet-based augmented reality (AR) game designed to enhance the social-emotional learning of children with autism. Utilizing advanced AI for generating personalized narratives and interactive activities, \name{} involves caregivers in a multi-phase learning process. In this engaging setup, children, assisted by their caregivers, navigate through customized stories and activities that foster emotion recognition and social interaction skills, integrating personal items and familiar environments to create an immersive and personalized experience.}
    \label{fig:teaser}
    \Description{}
\end{teaserfigure}
\maketitle

\input{texfiles/1-introduction}

\input{texfiles/2-relatedwork}

\input{texfiles/4-interfacedesign}
\input{texfiles/3-userscenario}

\input{texfiles/5-system}

\input{texfiles/6-evaluation}

\input{texfiles/7-results_quan_analysis}

\input{texfiles/8-results_qual_analysis}

\input{texfiles/9-discussion}

\input{texfiles/10-conclusion}

\begin{acks}
We thank all our participants for their time and valuable input. We would also like to thank our reviewers whose insightful comments have led to a great improvement in this paper. This work is supported in part by the NSERC Discovery Grant (2020-03966), Waterloo.AI's AI4SocialGood grant, and NSSFC Art Grant (22CG184).
\end{acks}

\bibliographystyle{ACM-Reference-Format}
\bibliography{bib/references.bib}

\newpage
\appendix
\begin{appendices}
\input{texfiles/Appendix-A_StoryExamples}
\input{texfiles/Appendix-B_ExpertsQuotations}
\input{texfiles/Appendix-C_Prompt}

\end{appendices}
\end{document}

%% file: texfiles/1-introduction.tex
\section{Introduction}
Autism spectrum disorder (ASD) is a complex neurodevelopmental disorder\footnote{The term ``autistic'' is used in this document to acknowledge that for many individuals, autism is an integral part of their identity. This identity-first language is preferred by a significant part of the autism community and is supported by recent academic advocacy efforts. We recognize that the preference for identity-first or person-first language can be personal and context-dependent. In cases where specific preferences are expressed, we endeavor to adapt our language accordingly to respect those choices~\cite{10.1145/3544548.3581341}.} characterized by impairments in social-emotional reciprocity, such as challenges with recognizing and expressing emotions and difficulties in understanding and responding to social interactions~\cite{DSM-5}. %
Such challenges are affecting approximately 1 in 100 children worldwide~\cite{world-health-organization}, with the number continuously growing~\cite{dinstein2023two, li2022global}. %
\emph{Social-emotional interventions} encompass various therapeutic approaches,
which have shown promise in supporting autistic children to better comprehend emotions~\cite{Golan2010}, enhancing facial emotion recognition skills~\cite{Tanaka2010}, and improving their responses to others' emotions~\cite{Bauminger2002}, thereby fostering more meaningful social interactions and relationships~\cite{Reichow2010}.

Social stories~\cite{gray_social_1993} offer a practical and effective method for social-emotional interventions to autistic children~\cite{Qi2018}, improving their skills like emotion recognition~\cite{Bader2006}. 
For instance, a social story can depict a scenario where a friend is crying due to sadness, and subsequently outline suitable reactions, aiding autistic children in understanding and reacting to such emotions~\cite{Bader2006}.
Nonetheless, it is still challenging to efficiently create high-quality social stories that fit autistic children's diverse social needs and personal preferences~\cite{Kokina2010}.
With the recent advancement in generative AI (GenAI), social stories have been generated with language that is easier for autistic children to understand, ensuring clarity and effectiveness~\cite{koegel_natural_1987, krantz_teaching_nodate}. 
While AI has the potential to greatly augment traditional social-emotional interventions, little research has been done to explore it in such a context. This has motivated our study to explore means of customizing social-emotional stories based on autistic children's living environments and personal needs, saving valuable time and resources.

In addition, social-emotional interventions require autistic children and their caregivers (i.e., their parents, family members, and teachers) to commit to long-term practices~\cite{Siller2002a}. 
Their involvement significantly affects the children's social-emotional skills and strengthens the bond between the caregivers and children~\cite{Vasalou2020Dfo, Frude2011Fsa, huang2022starrescue}. 
However, caregivers' active participation in interventions moderated by clinicians/ therapists for autistic children is hindered by financial constraints and limited access to services or resources~\cite{rogge_economic_2019, rios_exploring_2020}.
When outside of clinics such as at home or school, caregivers face the challenges of not equipping with enough professional knowledge or skills to guide their children in the interventions~\cite{narzisi2020handle}. 
Thus, there is a need for elaborately designing accessible and effective means to facilitate caregivers with getting involved in children's social-emotional development~\cite{Gengoux2019}.

Coupled with caregiver involvement, keeping children motivated in learning the emotions and paying attention to them is critical to generating more effective outcomes~\cite{Lyu2023}. 
However, traditional social-emotion delivering methods demonstrate expressions through cards-sorting or deck-sliding activities, which often require children to look at faces and eyes to interpret emotional expressions. However, these methods lack adequate features to engage autistic children since they naturally avoid looking at faces or eyes~\cite{Pelphrey2002Visual}.
Recent studies have demonstrated the effectiveness of interactive and immersive technologies, such as augmented reality (AR), in engaging autistic children.
By combining virtual and real-world environments, these technologies can help autistic children foster a sense of familiarity and safety~\cite{Lyu2023,chen_augmented_2016}.
Few works have investigated AR with GenAI to simultaneously promote penalization, caregiver involvement, and engagement in social-emotional story sharing and learning.

To fill in these gaps, we propose a novel tablet-based game, \name{}, that augments traditional social-emotional interventions~\cite{gray_social_1993, williams2012teaching} with personalized social stories, enhanced caregivers' participation, as well as interactive, fun, and immersive activities for autistic children. 
We aim to explore how \name{}, which integrates GenAI and AR and actively involves caregivers, impacts social-emotional interventions for autistic children in a collaborative manner.
The game encompasses five phases that are flexible to repeat, combine, or separate during a game session based on autistic children's needs (\autoref{fig:teaser}). 
1) \textit{Customization}: It begins with AI-infused customization that generates personalized social stories for a child. 
2) \textit{Comprehension}: It is followed by understanding emotions via storytelling with the caregiver. 
3) \textit{Observation \& Imitation}: Next, turn-taking exercises between the child and caregiver are dedicated to helping model and mimic facial expressions. 
4) \textit{Recognition}: Then, a dynamic AR activity prompts the child to recognize faces and emotions. 
5) \textit{Reinforcement}: Finally, the game concludes with reinforcement to solidify the learned concepts.
Built in these phases, \name{} also offers clear and concise instructions for caregivers, who do not need to be trained with specific intervention skills, to maximize the impact and outcomes of the intervention through active caregiver involvement. 
More specifically, we leverage the state-of-art AI model (\ie, GPT-4\footnote{\url{https://openai.com/}}) to generate social stories efficiently and customized based on a child's development needs and preferences.
Further, computer vision-based facial expression detection is incorporated with caregiver-child turn-taking to enhance children's motivation in observing and mimicking emotions.
Through a gaming process, \name{} incorporates AR to allow children to practice emotion recognition skills interactively in a familiar environment to boost their concentration and engagement.

\name{} was designed through an iterative process with five domain experts, including multiple stages from requirement gathering, prototype development, to initial evaluation. 
\rev{The co-design process emphasized the needs for adaptive storytelling with familiar scenarios, engaging visuals, and interactive, multimodal learning environments, along with a focus on clear emotional lessons and the use of pretend play to deepen understanding.
These insights derived the importance of personalization, caregiver involvement, and interactive engagement in design, guiding the development of \name{} with AI-generated narratives that integrate familiar objects and environments to address these needs.}
We further conducted a between-subjects controlled study with 24 autistic children and their caregivers in single-visit sessions at two special education centers to compare \name{} and a traditional intervention method as a baseline. 
The study provides both quantitative and qualitative findings about the \name{}'s usability, user experience of the system, and effectiveness of the proposed features.
\rev{Specifically, \name{} significantly improved emotion recognition skills among autistic children, and outperformed the baseline in usability and user experience.}
Our research also offers in-depth insights into the design and development of AI-infused AR-empowered tablet games as a practical and playful social-emotional intervention for autistic children.
\rev{
The integration of AI-enabled interactivity and tailored social stories involving familiar objects was seen as particularly effective in creating an engaging and immersive learning experience, contributing to better attention and sustained engagement.
The study highlighted the importance of real-world connections in AR-based activities, which helped maintain attention and promote a sense of achievement among children.}
In summary, our contributions in this paper include: 
\begin{itemize}
\item  An exploration of AI-infused approach that allows caregivers to be actively involved in the social-emotional development of their autistic children with tailored social stories and turn-taking.
\item  A novel AR-empowered tablet game to promote engagement in social-emotional learning for autistic children with interactive experiences. 
\item  Empirical insights into the design and development of relevant social-emotional learning tools from an iterative design process and a controlled user study.
\end{itemize}

%% file: texfiles/2-relatedwork.tex
\section{Background} \label{sec:relatedwork} 

\subsection{Social-Emotional Interventions for Individuals with ASD}
Social-emotional interventions target the core impairments of ASD, aiming for enhancing social and emotional well-being~\cite{nuske2023systematic}, including facial expression and emotions~\cite{jain2012interactive},  social communication and interaction~\cite{loiacono2018social, boyd2018vrsocial}. 
Among these interventions, social-emotional stories have been identified as particularly effective in fostering emotional understanding and improving social responsiveness in individuals with ASD~\cite{como2023scoping, ghanouni2019social}.
Moreover, computer-based interventions have emerged as a viable modality for augmenting social-emotional skills. 
Particularly when combined with supplementary tutoring, it can achieve outcomes comparable to those of traditional, in-person instruction, underscoring their potential as a valuable tool in the repertoire of ASD interventions~\cite{ramdoss2012computer}.
Noteworthy developments include an interactive game by Jain et al. that employs advanced computer vision to aid autistic children in learning facial expressions~\cite{jain2012interactive}. 
Similarly, Garcia-Garcia et al. have utilized recognition technologies to assist children in identifying and expressing emotions, further exemplifying the dynamic potential of digital tools in therapeutic settings~\cite{garcia-garcia_using_2022}.
Also, Boyd et al. have developed a collaborative game on iPads that strengthens social bonds among autistic children~\cite{boyd2015evaluating}. 
These works indicate the potential of delivering real-time, context-specific intervention to enhance the social-emotional development of autistic children. %

However, a key challenge for social-emotional intervention is in embedding the learning and reinforcement processes into everyday life for long-term commitment~\cite{slovak2016scaffolding}, complementing the in-class settings that the social-emotional interventions curricula are developed for~\cite{slovak2015teaching}.
The challenge highlights the potential for technology-based support to coach autistic children similarly to teachers and provide safe spaces for skill practice, allowing for relevant experiences and the possibility of failure~\cite{slovak2016scaffolding}.
In response, our work involves the design and development of a multi-phase intervention that augments in-class methods in an everyday setting to support social-emotional learning, including AI-generated social stories for emotional understanding, turn-taking for expression mimicking, and interactive activities for recognition practice.

\subsection{Role of Caregivers in Children's Therapeutic Processes}
Caregivers' active participation in the diagnostic and therapeutic process of children with developmental disorders is considered an important factor in the long-term struggle~\cite{Chaidi_Drigas_2020}.
For example, it is shown that parents' participation in ASD interventions is effective in improving parent-child relationships and bonding~\cite{hernandez2021parent}, social and communication skills~\cite{valeri_cooperative_2020}, motor skills~\cite{beaudoin_parent-mediated_2019}, and cognitive skills of autistic children~\cite{schertz_mediating_2018}. 
Parents can provide a consistent caregiving environment, integrating therapeutic practices into daily routines to enhance the child's learning and adaptation of social-emotional skills~\cite{sadeghi2021parent}.
Skills learned in therapeutic sessions are more likely to be generalized when caregivers reinforce them at home and in the community.
Technologies have been proposed to support parents' involvement in developing children's core developmental skills.

Notable examples include Marcu et al. using wearable cameras by parents to capture meaningful interactions, thereby promoting social engagement among autistic children~\cite{marcu2012parent}. 
Additionally, Venkatesh et al. introduced TOBY, an educational intervention delivered through iPads, empowers parents to initiate early intervention strategies aimed at optimizing their child's developmental outcomes~\cite{venkatesh2013toby}. 
Furthermore, Dunn et al. developed a parent-administered, tablet-assisted therapy for cognitive exercises tailored for autistic children~\cite{dunn2017tablet}. 

However, few studies in HCI have explored technological support for caregivers' involvement in interacting with autistic children to develop their social-emotional skills in daily contexts.
To this end, we introduce \name{}, offering opportunities for both verbal (\eg, storytelling) and non-verbal interactions (\eg, turn-taking), facilitated by a user-friendly interface that provides clear guidance and communication strategies to instruct caregivers through the intervention process.
Our approach aims to enrich the intervention landscape by embedding technologies like AI and AR %
thereby amplifying the parent-child relationship's impact as a foundational aspect of effective ASD intervention.

\subsection{Customization and Potential of AI for ASD Interventions}
Individuals with ASD exhibit a wide range of unique characteristics, due to the diversity in symptoms and their severity levels across the spectrum.
This diversity underscores the necessity of personalized approaches in gamified interventions for autistic children \cite{ritterfeld2009serious, susi2007serious}.
Customization is an important feature in serious games for individuals with ASD, as it creates the possibility to adapt the games to meet their preferences and needs~\cite{aruanno2018hololearn, bartoli2014designing, carlier2019using,loiacono2018social}.
Through a systematic review of 94 articles on serious games for autistic children, Carvalho et al.~\cite{de2023serious} identified four customization options:
1) \textit{different visual} (\eg, character's image) \textit{or audio} (\eg, volume level) \textit{elements to meet user preferences}~\cite{buzzi2019designing};
2) \textit{game activity parameters} (\eg, difficulty level)~\cite{aruanno2018hololearn}; 
3) \textit{support resources}  (\eg, positive or negative reinforcement)~\cite{carlier2019using};
and  
4) \textit{definition and creation of new activities}~\cite{crovari2019designing, rouhi2019emotify}.
The advent of GenAI a novel dimension offers personalized interventions beyond these categories~\cite{sabzevari2023artificial}, allowing children or caregivers to use real-life context in learning materials for optimal engagement and outcomes. 
The potential of GenAI in social-emotional interventions for ASD is vast but under-explored, presenting an opportunity to enhance evidence-based techniques via AI and customization.

Furthermore, AI has been integrated into emotion-based applications for various purposes, such as emotion therapy bots~\cite{wang2023training}, digital art co-creation~\cite{DU2024103139}, facial expression recognition~\cite{Washington2017-qc}, and emotional monitoring~\cite{bhargavi2023ai}.
This suggests a potential for AI to enhance social-emotional learning for ASD individuals with personalized support. 
For instance, Washington et al.~\cite{Washington2017-qc} utilize AI to present visual cues on emotional expression, assisting individuals with ASD in navigating social scenarios in real-life settings,
though caregiver involvement was not a focus.
Inspired by these studies, \name{} leverages AI to deliver personalized experiences and real-time feedback in social-emotional interventions for ASD, such as generating customized social stories and recognizing children's or caregivers' emotions, enhancing user engagement and enriching the intervention experience.

\subsection{AR Technologies for Supporting Autistic Children} 
Autistic children often face substantial challenges in navigating social environments~\cite{muller2008social}. 
However, they can significantly improve certain autism-related characteristics through computer-based communication~\cite{sturm2019participatory}. 
Augmented reality (AR) acts as a crucial bridge between tangible and virtual experiences, with autistic children showing remarkable openness to digital interventions~\cite{pennington2010computer,king2014ipad}. 
The incorporation of gamified elements within AR contexts has been particularly effective in engaging this demographic, suggesting a promising avenue for intervention~\cite{shane2008electronic,castellar2014improving}.

The application of AR in ASD interventions brings several advantages.
First, its portability stands out, with mobile devices such as smartphones and tablets serving as conduits for AR experiences, thus facilitating widespread access to virtual peers and assistive scenarios~\cite{tartaro2014accessing}. 
Moreover, AR offers a versatile and convenient option for treatment, especially suited for home environments, enhancing the accessibility of interventions. 
Crucially, AR's adaptability enables tailoring experiences to each child's interests, fostering deeper engagement.
Previous studies have delved into personalizing engagement by integrating children's preferred toys into the AR setting, enabling them to interact with these toys in a digital context~\cite{baio2018prevalence, dickey2006game, mercado2019developing}. 
Lyu et al.~\cite{Lyu2023} incorporated child-colored costumes for in-game characters in a neurofeedback training game, maximizing engagement and fostering emotional bonds. 
The personalized mechanism in previous studies not only enriches the gaming experience but also fosters emotional bonds between the children and their virtual counterparts.

Despite the established benefits of AR in ASD interventions, its effect on social-emotional learning with caregiver involvement is underexplored.
\name{} represents a first attempt to study these factors simultaneously, which synergizes AI and AR to enhance caregiver participation as well as provide immersive, context-aware, and personalized learning experiences.

%% file: texfiles/4-interfacedesign.tex
\section{Design of \name{}}

Our research proposes \name{}, a tablet-based game that enhances traditional social-emotional
interventions with tighter caregiver involvement, personalized social stories, as well as dynamic
and engaging activities.
Specifically, we developed \name{} through a co-design with five professionals in the domains of emotion learning, child education, and ASD. 

\subsection{Co-design Process} \label{sec:co-design}
We engaged five domain experts (E1-E5) with comprehensive backgrounds in neuroscience and education for autistic children. 
E1 is a university researcher with a focus on emotion-related neuroscience. %
E2 and E3 serve as educators at a specialized education center for autistic children. %
E4 and E5 are recognized experts in educational strategies for autistic children.  %
The experts have an average of 4 years of professional experience in their relative fields.  %
This design process also involved several autistic children and their caregivers who provided feedback on our initial ideas and prototypes.
Participants were voluntary during the co-design process, and confidentiality was ensured.
Spanning a year, this collaborative process facilitated a deep exchange of ideas and insights through questionnaires, interviews, co-design sessions, and remote discussions. 
We collected both qualitative and quantitative data to inform the design and development of \name{}.
Below, we detail the five main stages involved in this design process.

\textbf{Requirement Gathering (two months):}
This initial stage was aimed at acquiring a foundational understanding of ASD, grasping the nuances of social-emotional learning challenges faced by autistic children, and discerning principles for crafting effective interventions. 
A comprehensive literature review was conducted (\autoref{sec:relatedwork}) alongside a semi-structured interview with a caregiver of an autistic child, pinpointing the unique needs and characteristics of this demographic~\cite{aresti-bartolome_technologies_2014}. 
Our approach was informed by the investigation into existing technologies~\cite{garcia-garcia_using_2022} and their role in enhancing social-emotional skills and the analysis of HCI research on design challenges and opportunities~\cite{hailpern_encouraging_2007}. 
We also carried out two semi-structured interviews with E4 and E5, which led us to choose an easy-setup mobile platform to access AI and AR. 
We analyzed the interview transcripts and generated several requirements (\autoref{sec:design-insights}) for designing caregiver-involvement games to promote engagement and intervention outcomes for autistic children.

\textbf{Early Concept Design (two months):}
Informed by the distilled insights into how social-emotional skills are taught in schools, we envisioned a game encompassing five phases, each incorporating elements deemed effective for social-emotional interventions tailored to autistic children. 
Based on the requirements and our literature survey, we solidified the design principles (\autoref{sec:design_principles}) for developing the intervention, leading to the conception of AI-generated social stories and a gamified strategy with the integration of AI and AR. 
Further, to enable story generation with AI, we carefully examined guidelines from the literature~\cite{reynhout_social_2006, gray_social_1993}, coupled with suggestions from three of the experts (E1-E3).
We then fine-tuned OpenAI's GPT-3.5/4 to generate stories for seven basic emotions (\ie, happiness, surprise, sadness, fear, disgust, anger, and neutral), which were later assessed by our experts with an online survey (\autoref{sec:emotional-stories}).

\textbf{System Development (five months):}
Through frequent consultation with E1-3, we deliberately designed five characters and seven emotional faces for each to fit the needs of autistic children (\autoref{sec:game-characters}).
We developed a working prototype of \name{} as a tablet application, incorporating both the enhanced model and the emotionally expressive characters (\autoref{sec:scenario}).
We also strategically designed features to aid in the learning process (\autoref{sec:ar-activity} and \autoref{sec:ai-interaction}), such as interactive turn-taking, emotion-based AR activities, and customizable stories. 
We integrated clear in-game guidance and communication strategies for caregivers, aiming at facilitating effective caregiver-child interaction during the game sessions.
This cooperative approach ensured that each feature of \name{} was meticulously crafted to support the nuanced social-emotional learning needs of autistic children.

\textbf{Initial Assessment (one month):}
Using the working prototype, we conducted a pilot study with two autistic children, accompanied by their caregivers at a special education center. 
We observed the children as they engaged with \name{}, then asked the caregivers to complete a questionnaire about their experience and observation, followed by a semi-structured interview.
Overall, they appreciated the game and mentioned that the children were notably more interested in the interactive game than traditional means.
We refined the game based on the feedback, which included incorporating more exaggerated facial expressions and opting for less colorful clothing on characters to better focus children's attention on emotional faces. 

\textbf{Evaluation (two months):}
Using the refined prototype, we conducted a between-subject controlled study involving 24 autistic children and their caregivers, to gather empirical findings from both qualitative (\ie, interviews) and quantitative (\ie, Likert scale rating) data (\autoref{sec:evaluation} and \autoref{sec:results}).
We compared \name{} with a baseline using traditional slides-based learning.
The findings highlight \name{}'s ability to enhance engagement and improve learning experiences and outcomes for autistic children, signifying a substantial leap forward from conventional practices.
The findings also include a set of design implications that could inform future research in this area (\autoref{sec:disscussion}).

\begin{figure}
    \centering
    \includegraphics[width=1.\linewidth]{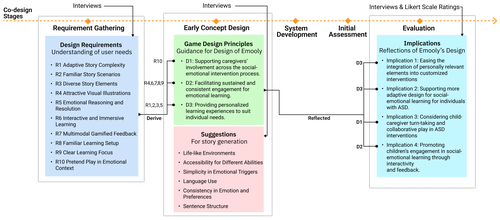}
    \vspace{-8mm}
    \caption{During the co-design process, qualitative and/or quantitative data informed each stage from the initial requirement gathering to the final evaluation. This process demonstrates how design requirements, design principles, and story generation suggestions are related to guide the design and development of \name{}, ultimately resulting in implications that shed light on future research.
    }
    \label{fig:relationship}
    \Description{}
\end{figure}

Through the process, we gained the design requirements (\autoref{sec:design-insights}), design principles (\autoref{sec:design_principles}), suggestions (\autoref{sec:evaluation}), and implications (\autoref{sec:disscussion}) that served different roles in our design and evaluation.
\autoref{fig:relationship} organizes them and highlights their interrelationships. 
The design requirements (R1-R10) provide a foundational understanding of the need for implementing effective interventions for autistic children.
From these requirements, we derived specific design principles (D1-D3) to guide the design and development of \name{}. 
Experts' suggestions offer practical strategies for creating social-emotional stories for social-emotional learning for autistic children. 
Finally, our final evaluation provides the implications reflecting \name{} in educational settings and its potential impact on social-emotional learning for autistic children.

\subsection{Design Requirements}  \label{sec:design-insights}
Based on our co-design process with the experts, we derived the following requirements that shed light on the development of social-emotional games for autistic children. 
We employed an emergent coding approach in qualitative analyses to identify common themes.
Experts' quotations were coded and categorized in \autoref{appendix:experts-quotations} to support each design requirement.

\begin{enumerate}[leftmargin=8mm, label=\textbf{I\arabic*}]
\item[\textbf{R1}] \textbf{Adaptive Story Complexity:} Tailor stories to align with the social-emotional skills and cognitive abilities of children in a spectrum of autism. This is essential to keep children engaged and progress over time. 
\item[\textbf{R2}] \textbf{Familiar Story Scenarios:} Incorporate life-like scenarios that autistic children might find familiar to facilitate the understanding of complex emotions. 
By embedding emotional experiences in recognizable contexts, children can navigate and internalize these emotions more effectively and motivated.
\item[\textbf{R3}] \textbf{Diverse Story Elements:}
Different scenarios, characters, and stories help children learn through the emotional reflections of different characters in the stories.
This diversity is essential for fostering a comprehensive emotional understanding through varied and relatable narratives.
\item[\textbf{R4}] \textbf{Attractive Visual Illustrations:} Utilize visually engaging elements within the learning process as essential stimuli for capturing the attention of autistic children and enhancing their learning outcomes.
\item[\textbf{R5}] \textbf{Emotional Reasoning and Resolution:} Facilitate understanding of emotions by teaching their causes, effects, consequences, and coping strategies through story-based exploration to enrich comprehension, while simulated experiences prompt deeper emotional insights.
\item[\textbf{R6}] \textbf{Interactive and Immersive Learning:} Provide a more dynamic and situational environment. This can offer children a deeper understanding of the learned emotions and enable them to practice and reflect more effectively.
\item[\textbf{R7}] \textbf{Multimodal Gamified Feedback:} Prioritize multimodal channels (\eg, visual and auditory) in a gaming context for autistic children's engagement, integrating feedback dynamically and smoothly to avoid distraction. 
\item[\textbf{R8}] \textbf{Familiar Learning Setup:} Choosing familiar and relevant scenarios to teach emotional expressions, making emotional learning more effective and engaging by reducing external interferences and spotlighting crucial educational content.
\item[\textbf{R9}] \textbf{Clear Learning Focus:} Making key learning elements more prominent can effectively enhance focus and comprehension. The approach of accentuating human facial expressions within the learning content significantly aids in capturing attention. 
\item[\textbf{R10}] \textbf{Pretend Play in Emotional Context:} 
Having children mimic emotions with contextual scenarios to learn emotions as pretend play. Although mimicking expressions may not always evoke genuine feelings, using pretend play as a guide can facilitate deeper emotional understanding. 
\end{enumerate}

\subsection{Design Principles} \label{sec:design_principles}
Based on the obtained requirements and the literature, we consolidated the following design principles for the social-emotional game.
The final prototype of \name{} can be seen as an embodiment of these principles, allowing us to investigate the approach to integrating AR and AI in a mobile setting. %

\textbf{D1: Supporting caregivers' involvement during children's} social-emotional intervention.
Caregivers' active participation in the intervention process of autistic children is considered by experts to be an important factor in the long-term struggle~\cite{Chaidi_Drigas_2020}.
Thus, the system should be designed to promote participation from both sides to ensure the effectiveness of the intervention~\cite{Wilcox2011Participation}, encourage communication and interactions for learning outcomes, and concurrently strengthen the bond between the child and their caregiver~\cite{theofanopoulou2022exploring}.
In particular, the overall social-emotional intervention should be split into manageable phases that are easy to follow~\cite{park2012framework}, and promote learning through caregiver-child interactivity such as turn-taking (R10).
Meanwhile, the system should provide caregivers with expert-validated communication strategies and clear guides, ensuring an environment conducive to participation and learning~\cite{alli2015parents}. 
Additionally, the system should be designed to be user-friendly, ensuring it is intuitive and easy to use, allowing children and caregivers to engage with the system without additional stress or complexity.

\textbf{D2: Facilitating sustained and consistent engagement for emotional learning.}
It is essential to maintain the engagement of autistic children and promote their continuous involvement in a gamified manner~\cite{Whyte2015-fs}. 
Incorporating a variety of tasks that intersperse various activities such as AR-based~\cite{koegel2003empirically} as well as integrating different interactive techniques and feedback have proven effective in enhancing the interest, excitement, and joy in individuals with ASD~\cite{boucenna2014interactive, de2023serious} (R6, R7).
Further, the learning experiences should be rooted in the environments and situations that children are familiar with (R8).
This dynamic approach needs to be combined with emphasizing expressions to draw attention (R9) and incorporated visual illustrations (R4), allowing children to interact with them, foster comprehension, maintain attention, and enhance their engagement.

\textbf{D3: Providing personalized learning experiences to suit individual needs.}
ASD individuals are highly unique since there are different symptoms and severity levels within the spectrum.
Empowering children to take the lead and make choices allows them to craft stories and activities that resonate with their interests and are set within familiar environments~\cite{quintero2010something}. 
Designing games with the capability for such customization requires a flexible approach, such as allowing for the modification of storylines, characters, and challenges to suit the child's skill level and interests, thereby optimizing the learning experience~\cite{Whyte2015-fs}.
In particular, tailored social stories (R5) that have various difficulty levels (R1), social scenarios (R2), and story elements (R3) should be carefully considered in the intervention design.

%% file: texfiles/3-userscenario.tex
\section{\name{} Walk-through}  \label{sec:scenario}

\begin{figure}
    \centering
    \includegraphics[width=1.\linewidth]{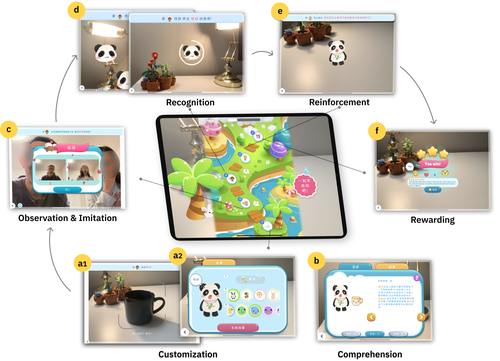}
    \vspace{-8mm}
    \caption{Workflow of \name{}: (a1,2) Customization: The child personalizes the social-emotional story by combining real-world objects into AI-generated narratives with a selected character and target emotion.
    (b) Comprehension: The child reads and understands the generated story enhanced with visual illustrations and follow-up questions, accompanied by the caregiver. 
    (c) Observation \& Imitation: The child and the caregiver perform turn-taking activities to facilitate the practice and replication of emotional expressions.
    (d) Recognition: The child is engaged in a dynamic AR activity to identify the target emotion from various emotional expressions overlayed in the real-world environment. 
    (e) Reinforcement: The child and the caregiver conclude the session with reflective questions that bridge the target emotion with real-life experiences.
    (f) Communication tips are available at various places of the user interface during the session.
    }
    \label{fig:UI-flow}
    \Description{}
\end{figure}

Before going into the detailed implementation of \name{}, we outline a simple scenario to walk through the use of \name{} which enriches social-emotional understanding and recognition through immersive and interactive stages. 
Mia is a six-year-old girl with autism; accompanied by her mother (D1), Mia engages with \name{} by following milestones on a map, each corresponding to a phase in the intervention.
The goal of the game is to complete the tasks in each milestone successfully and collect ``hearts'' along the way, which will be presented as rewards at the end of the session.
\autoref{fig:UI-flow} demonstrates the whole process described in the following.

\textbf{Customization:}
The journey commences as Mia, equipped with an iPad, embarks on personalizing her gameplay experience (D3). 
Guided by on-screen text and icons, she begins by capturing an image of her cherished mug, a beloved gift from her last birthday, integrating this personal element into \name{}'s narrative (R5). 
This phase is instrumental in engaging Mia by crafting a connection between her real-world environment and the game's digital storyline via personalization. 
Mia's mom guides her through the game, following some text instructions; Mia navigates and clicks on icons representing characters and emotions.
She selects the character ``Panda'' and the emotion ``Surprised,'' because a panda resonates with her (R3). 
This fosters an emotional bond with the content of the narratives.

\textbf{Comprehension:}
After the customization, \name{} generates a personalized narrative that incorporates Mia's selections, tailored to her level of social-emotional understanding (D3). 
The generated story says \qt{Patty got a special mug as a gift during lunchtime in kindergarten. When he put cold water in it, the mug changed colors like a rainbow! Patty was so surprised. He showed his friends, and they all thought it was really cool! Patty loves sharing discoveries with his friends, which makes him more excited.}
The narrative can also be adjusted for complexity based on input from Mia's mom (R1), striking a balance between challenge and accessibility. 
Clicking the button labeled ``More Difficult,'' Mia's mom wants to generate a story that challenges Mia while maintaining the character and object she selected.
Mia earns three hearts by participating in the storytelling session and answering the follow-up questions.
This phase encourages exploration of the story and its social-emotional cues (R4), with follow-up questions that can be discussed with her mom (D1), aimed at building Mia's comprehension of the social-emotional landscape presented (D2).

\textbf{Observation \& Imitation:}
In this phase, Mia's mom first demonstrates the emotional expression introduced in the narrative, and then invites Mia to mimic it (R10), incorporating a turn-taking mechanism that can be kept going.
Throughout this phase, they follow the visual cues and clear instructions that indicate whose turn it is and who should be waiting.
Mia's mom first poses a surprised face in front of the camera and Mia mimics next. 
This mimics real-life turn-taking situations, providing practical experience in a controlled environment.
Supported by AI-driven visual feedback, this turn-taking is essential to encouraging Mia to practice and refine her emotional expression skills, thereby enhancing her ability to recognize and replicate emotions accurately (D1).
Mia also receives hearts for her efforts in participating in the activity to reinforce this exercise.

\textbf{Recognition:}
In this phase, the game transitions to an AR activity, designed to focus on emotion recognition in an interactive and immersive way (R6, R9).
Here, Mia is tasked with identifying various emotional expressions of the panda through the tablet’s camera, which overlays these expressions onto her real-world environment (R8). 
As the expressions pop up on her screen, each showcasing a different emotion, Mia must correctly recognize and match these emotions to a target emotion presented in the game with positive feedback (R7). 
For each correct match, she receives positive feedback and is rewarded with hearts at the end based on her performance.
This exercise not only solidifies Mia's understanding of different emotional expressions by placing them in a tangible context but also enhances her engagement through the use of AR, making the learning experience both immersive and enjoyable (D2).

\textbf{Reinforcement:}
Concluding the journey, Mia engages with ``recap'' questions (\eg, \qt{What would you do when you discover something surprised?}) prompting reflection and personal experience~\cite{Wong1979Increasing}, also encouraging more discussion with her mom (D1). 
The question serves to bridge the story's emotional themes with Mia's own life (R2).
This phase aims to enrich Mia's social-emotional comprehension by drawing parallels between Patty's surprise with friends and similar feelings Mia might have encountered. 
The format of these open-ended questions fosters a genuine dialogue about emotions and experiences, thereby enhancing Mia's ability to articulate her thoughts and feelings, and furthering her social-emotional development in the long term (D3).

\textbf{Rewarding:}
Mia accumulates hearts as positive feedback for her participation and performance in the above phases, and ultimately receives brightening stars as a final reward on the screen. 
The stars are accompanied by animated icons and uplifting music, creating a festive atmosphere that celebrates her achievements.
For example, a smiley face provides encouragement, and a ``You win!'' banner enhances the sense of accomplishment.
Moreover, Mia's mom receives tips to improve interaction and communication, such as using the ``Your Turn'' strategy during turn-taking games to help Mia understand the concept of taking turns.

\begin{figure}
    \centering
    \includegraphics[width=1.\linewidth]{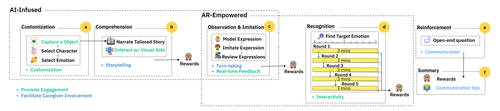}
    \vspace{-8mm}
    \caption{Beginning with AI-infused customization, the game progresses through comprehension with visual storytelling, leverages AR for observational learning and emotion recognition, and concludes with reinforcement techniques to enhance communication skills. \name{}'s design aims to promote active engagement and encourages caregiver participation, which is essential for the learning outcomes.}
    \label{fig:procedure}
    \Description{}
\end{figure}

Throughout her journey with \name{}, Mia enjoys personalized narratives and interactive activities designed to boost her emotion recognition and social interaction skills, alongside immersive exercises for deeper emotional understanding (D2, D3).
Additionally, her mom's active involvement in the process enhances Mia's learning experience, and when appropriate, her mom is supported with tips and guidance popped up by \name{} (D1), which makes the essential support to Mia's emotional development easier, engaging, and fun. 
A conceptual level summary of the whole procedure of interacting with \name{} is shown in \autoref{fig:procedure}.

%% file: texfiles/5-system.tex
\section{System Implementation}
In this section, we detail the implementation of key components in our system.
We developed \name{} as a mobile AR application on iPadOS with Unity for ubiquitous access. 
It can be deployed on multiple mobile platforms, while a tablet is recommended for ease of interaction.

\subsection{Game Characters and Emotional Faces Design} \label{sec:game-characters}

\begin{figure}[tb!]
    \centering
    \includegraphics[width=1\linewidth,trim=0 10 0 10,clip]{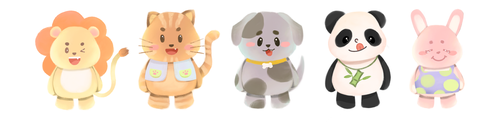}
    \vspace{-10mm}
    \caption{Cartoon characters in \name{}.}
    \label{fig:characters}
    \Description{}
\end{figure}

In the customization phase (\autoref{fig:procedure}-a), \name{} allows the child to choose from various emotional characters (\autoref{fig:characters}).
We carefully crafted five cartoon characters, and these selected characters remain constant throughout the entirety of the game session, providing a personalized and cohesive gaming experience.
Because autistic children may have difficulty recognizing complex emotions, we choose seven basic emotions (\autoref{tab:emotional-faces}), which are universal and can be recognized purely as emotions without the need to attribute a belief to the person~\cite{Golzari2015}.
Moreover, previous work emphasizes the importance of breaking down complex emotions into simpler components and basic emotions as a foundation for understanding more nuanced emotions for learning effectiveness~\cite{Golan2006,BaronCohen2001}.

\input{tables/cartoon-faces}

We use cartoon characters because it is posited that cartoon faces could serve as an engaging medium for emotional education given the affinity many autistic children have for cartoons~\cite{rosset2008typical}. 
This preference is further supported by evidence suggesting that they often favor simplified facial expressions for ease of interpretation~\cite{Ricks2010Tac, Deriso2012EMA}. 
Acknowledging the challenges these children might face in navigating real-world social interactions, we opted to utilize cartoon representations of animals with simplified human expressions in the game. 
This design choice focuses on key facial features---such as eyes, mouth, and eyebrows---to vividly portray a spectrum of emotions, according to the FACS (Facial Action Coding System)~\cite{Ekman1978FAC}.
The simplification to non-human cartoons not only makes the emotional expressions clearer but also reduces the social complexity often associated with human faces. 
This approach aims to facilitate a more accessible and less pressured learning environment for autistic children, allowing them to understand and learn about emotions more comfortably~\cite{rosset2008typical}.

\subsection{Social-Emotional Story Generation} \label{sec:emotional-stories}
Based on personalized settings, 
\name{} generates a narrative aiming to enhance the child's social-emotional understanding (\autoref{fig:procedure}-b). 
\name{} is designed to be user-friendly for both children and caregivers. 
To generate a story, children first use the camera of a tablet to scan objects that interest them, which caregivers can prepare in advance to have a variety of meaningful choices. 
Next, children select emotions from a system-provided list, with caregivers' assistance if needed. 
Caregivers can then adjust the story difficulty to suit the child's abilities and needs. 
The generated story is displayed as text with visual illustrations, providing an engaging experience to enhance children's comprehension and social-emotional learning.

\subsubsection{Generative AI Model}
The narrative generation thus considers three key aspects. First, to elevate the contextual relevance of the stories, \name{} leverages OpenCV\footnote{\url{https://opencv.org/}} to detect objects in the camera scenes of the child's tablet, allowing for scanning objects from their environment to weave them into the storyline (\autoref{fig:UI-flow}-a1). 
We encourage children to incorporate objects of personal significance into the game, promoting a sense of ownership and active participation.
This approach bridges the gap between the tangible world and the narrative sphere, enhancing engagement, motivation, and educational outcomes by grounding the learning experience in the children's immediate reality.

Second, based on the selected object, character and emotion, the generation integrates a variety of sentence types---descriptive, perspective, directive, control, affirmative, and cooperative---each chosen for its unique contribution to the narrative's effectiveness, following the writing guideline by Gray~\cite{gray_social_1993} (\autoref{fig:UI-flow}-a2). 
To enrich our stories with professional insights, we combined these structured sentence types with the following experts' suggestions, distilled during our co-design, in the prompts of GPT-4 for a high-quality generation:
\begin{itemize} 
 \item \textbf{Life-like Environments:} Including items and settings that closely resemble real life in the stories can benefit children's life experiences for generalization, which highlights the importance of incorporating realistic elements into stories.
 \item \textbf{Accessibility for Different Abilities:} Stories should be adaptable for children with varying degrees of autism, including those without speech functions, who may require more guidance through language and visual cues.
 \item \textbf{Simplicity in Emotional Triggers:} The causes of emotions in stories should be straightforward and instinctual, such as disgust from dirtiness or happiness from recovering a lost item, reflecting basic human nature.
 \item \textbf{Language Use:} The vocabulary should be simple, with concrete nouns for lower levels and possibly abstract terms for higher levels, but always kept understandable. 
 \item \textbf{Consistency in Emotion and Preferences:} In stories of low difficulty, maintaining consistent attitudes and preferences is advised to avoid confusing children with different emotional responses.
 \item \textbf{Sentence Structure:} Longer sentences should be broken down to aid children in understanding and thinking more clearly.
\end{itemize}

Third, the story generation in \name{} presents different levels of difficulty, enabling caregivers to select the most appropriate level for their child (\autoref{fig:UI-flow}-b).
To have varying difficulties, we instructed the AI to generate a narrative that is either harder or simpler than the current one based on the caregiver's input, where they can choose ``more difficult'' or ``easier'' to regenerate the story with more complex or simpler sentences, language, and situations. 
The complexity level is shaped by the narrative content, language used, and word choice, based on the insights from our experts.
In principle, a more difficult story refers to more plot twists, character interactions, or more complex emotional expressions, which involves a more complicated process of emotional management, requiring a greater understanding and application of social skills.

Overall, the model development went through a rigorous design process as outlined in our co-design (\autoref{sec:co-design}).
In specific, the process consisted of the following key steps, with the involvement of our experts:
1) \textit{Prompt Creation:} Crafting appropriate prompts with initial experimentation;
2) \textit{Initial Generation:} Producing a set of sample stories from the model with the prompts;
3) \textit{Expert Review:} Having experts review the stories for appropriateness and share feedback;
4) \textit{Feedback Incorporation:} Refining the prompts based on experts' feedback in multiple iterations;
5) \textit{Final Approval:} Finalizing the prompts to generate stories for evaluation (\autoref{sec:model_eval}).

Examples of generated stories are presented in \autoref{appendix:stories-examples}.
GPT-4 was provided with specific prompts and the key aspects, including the selected object, character, emotion, and difficulty level, ensuring it is relevant and engaging for the child.
The prompts are also provided in Appendix~\ref{appendix:prompt}.

\subsubsection{Model Evaluation} \label{sec:model_eval}
To evaluate the GenAI model in our system, we used it to generate 56 stories, 8 for each emotion, and later invited our experts (E1-E3) to assess the quality of the stories using a 7-point Likert scale (1-very inappropriate to 7-very appropriate). This method was chosen to provide a quantitative assessment of the perceived quality and suitability of AI-generated stories. We computed the median rating for each emotion category as well as for each expert (\autoref{tab:stories-rates}).
The overall median rating across all emotions and experts was approximately 5.5 out of 7, indicating a generally favourable assessment of the stories' suitability for the social-emotional intervention.

The ratings revealed that positive emotions (\eg, happiness) and strongly felt negative emotions (\eg, anger, surprise, and disgust) received favorable feedback (with median ratings above 5), which indicates the AI-generated narratives' capability to craft engaging and emotionally-resonant content for ASD social-emotional intervention.
However, the challenge of accurately eliciting fear was spotlighted by a lower rating, especially from E1.
The expert pointed out that the story scenarios, specifically, the loss of belongings, are not naturally inducing fear.
This observation stems from an understanding that fear typically arises from facing or anticipating danger without the ability to cope, characterized by feelings of shock and crisis. 
The variation of ratings underscores the complexity of emotional experiences.
E3 further emphasized the need for straightforward reasons for arousing emotions within stories. 
This suggests that for AI-generated narratives to be more effective in eliciting specific emotions like fear, they must not only include clear and relatable triggers but also account for the diverse ways individuals perceive and react to different scenarios.

\begin{table*}[tb!]
\small
\centering
\caption{Median ratings on generated emotional stories for different emotional categories and experts, on a scale from 1-very inappropriate to 7-very appropriate.}
\vspace{-3mm}
\begin{tabular}{lrrrr}
 \toprule
 \textbf{Emotion} & \textbf{E1} & \textbf{E2}  & \textbf{E3} & \textbf{Overall} \\ 
 \midrule
Anger & 7 & 5 & 5 & 5\\ 
Happiness & 7 & 6 & 4.5 & 6 \\
Sadness& 5.5  & 6 & 5 &  5.5 \\ 
Disgust&  7  & 5 & 5 & 5\\ 
Fear &  3.5 & 5 & 4.5 & 4.5\\  
Surprise & 5.5 & 6 & 5  & 5.5  \\ 
Neutral & 5.5 & 5.5 & 5 & 5.5  \\ 
\midrule
\textbf{Overall}   & 5.5 &	5.5 &	5 &	5.5\\
\bottomrule
\label{tab:stories-rates}
\end{tabular}
\end{table*}

\subsubsection{Other Story Elements}
Alongside the narrative, \name{} incorporates visually engaging elements to enhance the storytelling experience, making it more interactive and appealing.
We illustrate the selected object alongside the emotional characters next to the generated story, enabling children to engage through touchscreen interactions.
Upon tapping, the characters animate in ways that reflect their displayed emotions. 
For instance, a character with anger will show a pulsating animation conveying the intensity of this emotion.
Following E3's recommendation, emotion-related words are highlighted to draw attention.
We also use simple backgrounds for the visual elements, minimizing distractions for children with autism, and ensuring a clearer and more focused interaction.
Additionally, we instructed the AI to generate five follow-up questions along with potential answers for each story, for the caregiver to communicate with children to better understand the emotion in the story and the social scenario. 
Similarly, a ``recap'' question is generated by the AI to show in the reinforcement phase to help deepen their understanding by reviewing the emotions and thoughts children may experience in similar situations (\autoref{fig:UI-flow}-e).

\subsection{AI-empowered Interactive Turn-taking } \label{sec:ai-interaction}

\begin{figure}[tb!]
    \centering
    \includegraphics[width=0.24\linewidth,trim=0 10 0 10,clip]{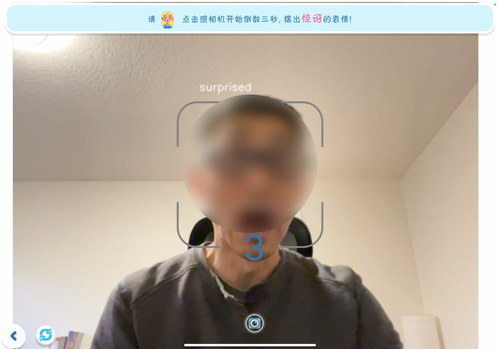}
     \includegraphics[width=0.24\linewidth,trim=0 10 0 10,clip]{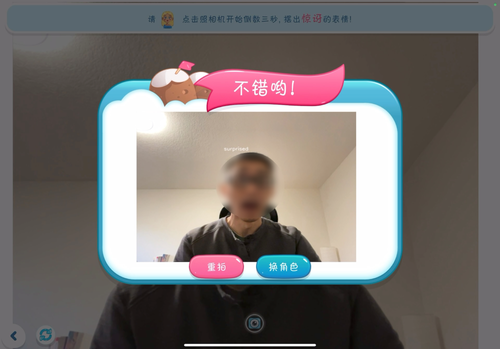}
     \includegraphics[width=0.24\linewidth,trim=0 10 0 10,clip]{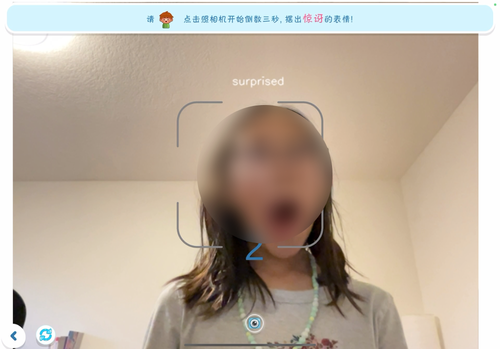}
     \includegraphics[width=0.24\linewidth,trim=0 10 0 10,clip]{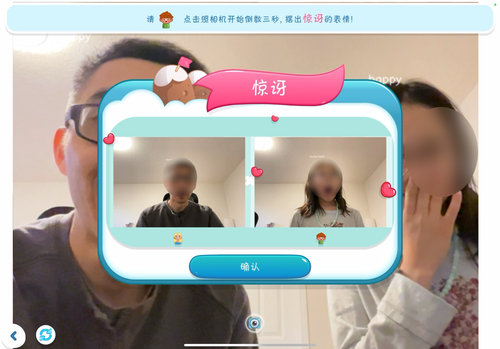}
    \vspace{-3mm}
    \caption{In the observation \& imitation phase, the caregiver first presents a facial expression captured and recognized by \name{}, setting the stage for the child to mimic in a turn-taking manner. After the child presents their facial expression, \name{} puts both images side-by-side to highlight the turn-taking success and reinforce learning.}
    \label{fig:mimic-activity}
    \Description{}
\end{figure}

In the observation \& imitation phase (\autoref{fig:procedure}-c), \name{}'s turn-taking mechanism between autistic children and their caregivers is enriched by computer vision techniques. 
Here, caregivers model specific facial expressions tied to the stories, prompting children to mimic these expressions with precision (\autoref{fig:mimic-activity}). 
We leverage facial recognition to evaluate both the caregiver's modeling and the child's mimicry in real-time and provide instant feedback. 
In specific, once the child or caregiver aligns their face within a frame on the screen, \name{} utilizes OpenCV to recognize and classify the facial expression.
This process involves detecting facial landmarks and analyzing them to infer the displayed emotion, which is then labeled and shown adjacent to the live image. 
At the end of this turn-taking, \name{} places the captured images of caregivers' and children's faces side by side, which not only encourages replication of expressions but also bolsters their understanding of emotional cues (\autoref{fig:UI-flow}-c). 
Concurrently, the AI-supported turn-taking feature reinforces social cue learning through repeated practice and feedback.
In addition, the photos would be automatically added to the photo gallery for caregivers or children to review the progress.
Therapists can also utilize these histories to explore, analyze, and communicate the progress of children, which can be difficult to accomplish otherwise~\cite{gelsomini2016affordable,zaman2014usabilty}.

\subsection{Dynamic Augmented Reality Activity} \label{sec:ar-activity}

\begin{figure}[tb!]
    \centering
    \includegraphics[width=0.24\linewidth,trim=0 10 0 10,clip]{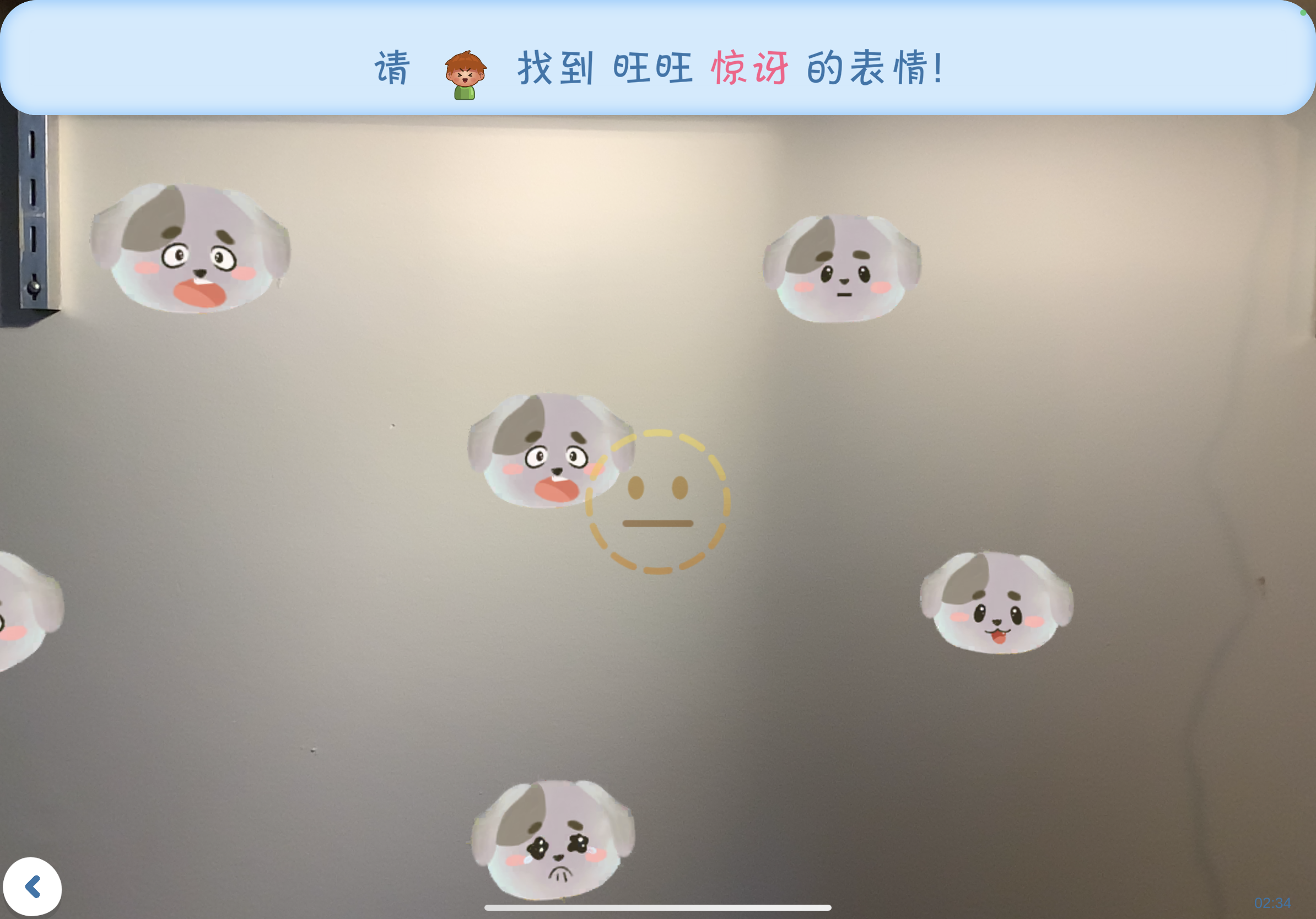}
    \includegraphics[width=0.24\linewidth,trim=0 10 0 10,clip]{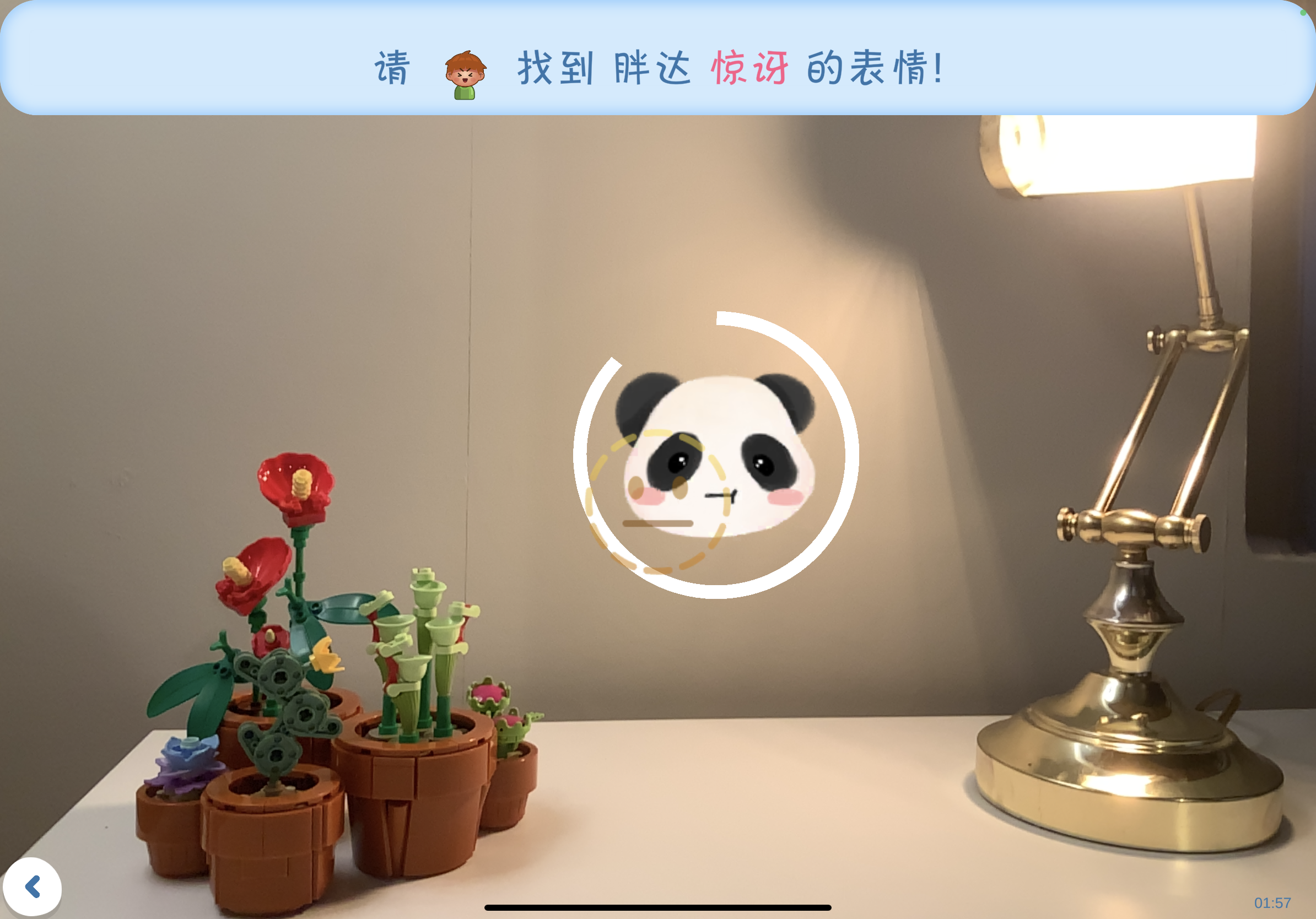}
    \includegraphics[width=0.24\linewidth,trim=0 10 0 10,clip]{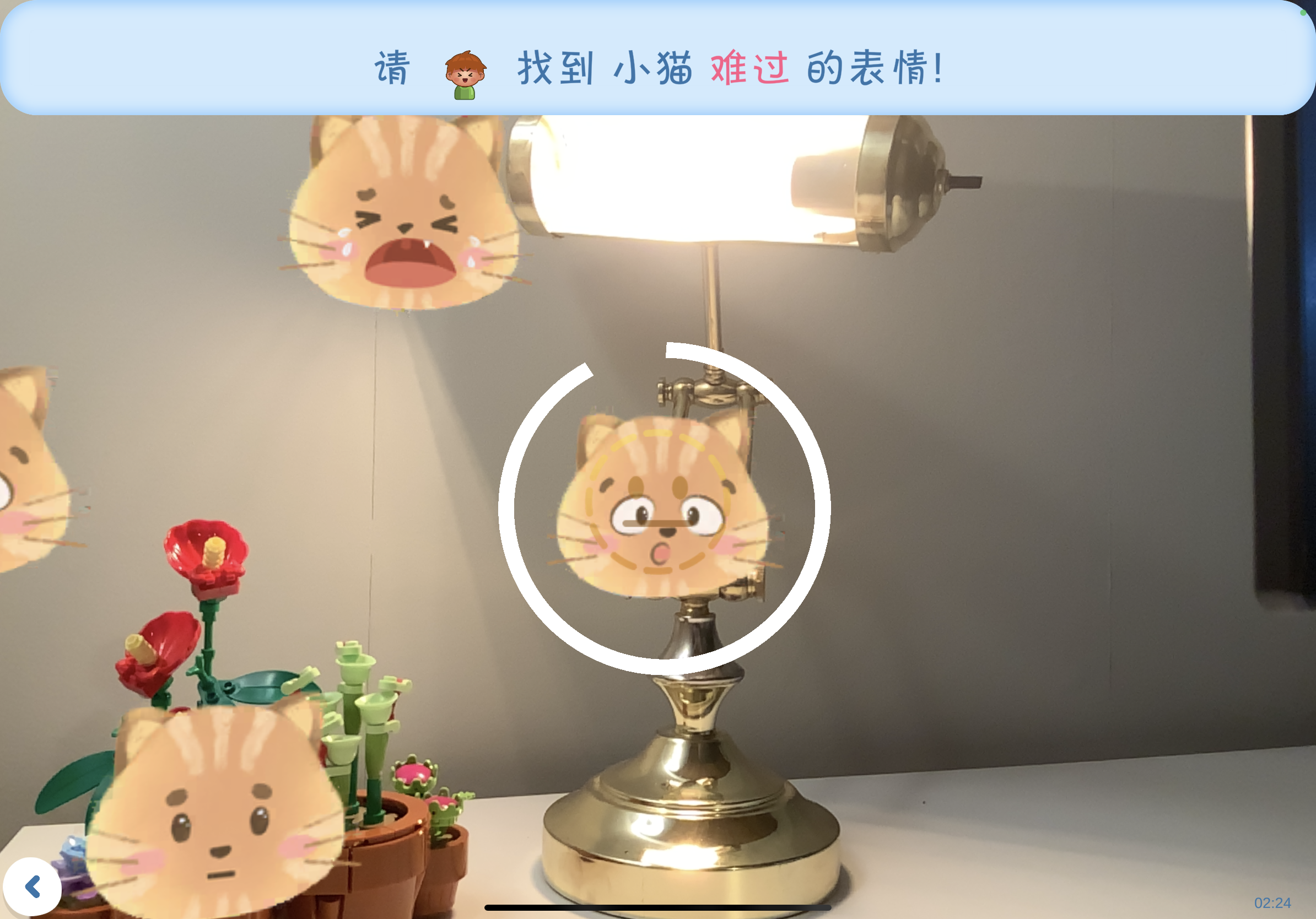}
     \includegraphics[width=0.24\linewidth,trim=0 10 0 10,clip]{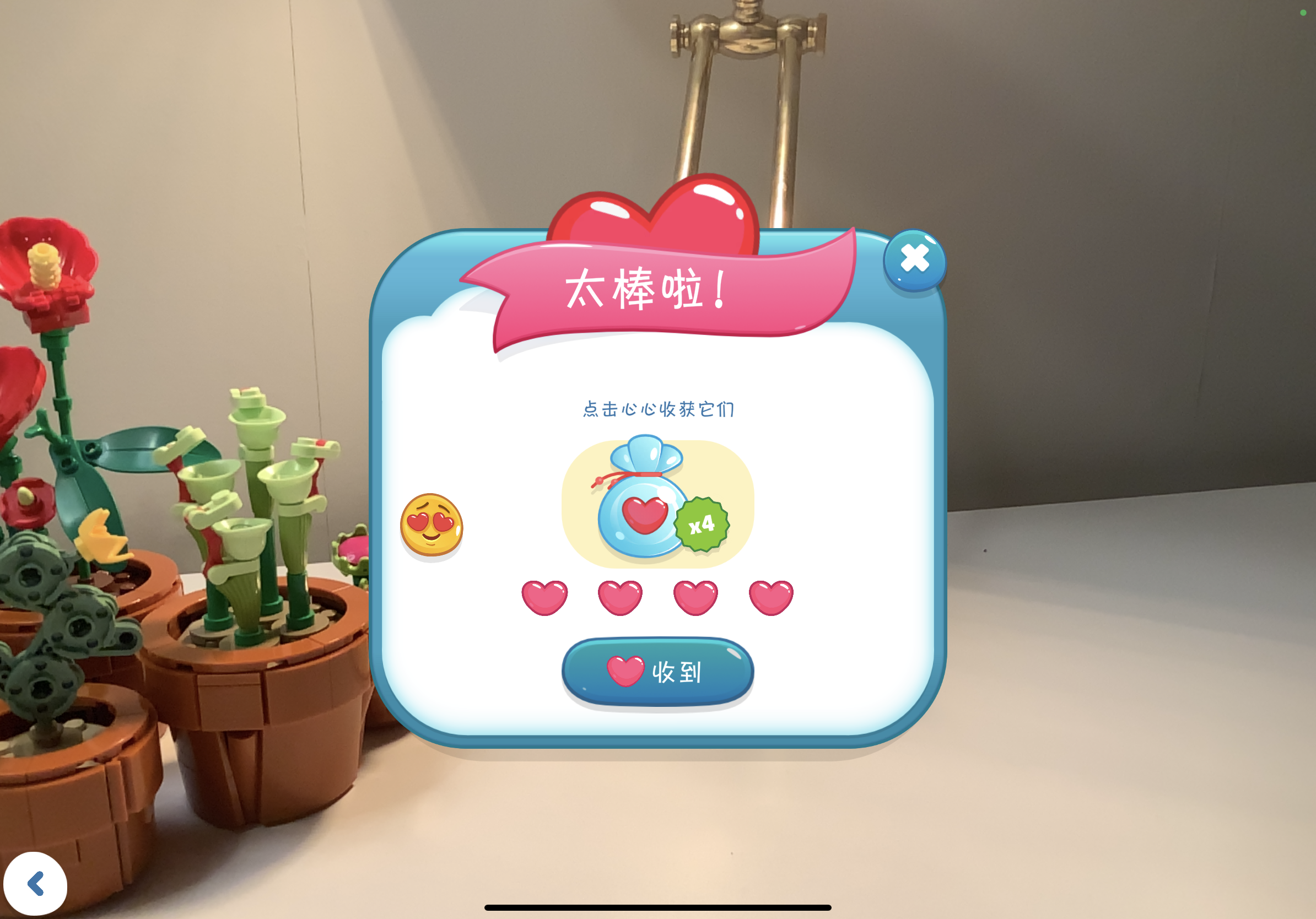}
    \vspace{-3mm}
    \caption{In the recognition phase, \name{} displays a variety of digital facial expressions around a virtual sphere from the child's perspective, introducing different emotions they need to recognize. The child aligns a facial expression in the center screen area to ``capture'' a target emotion. Correct captures result in cheerful animations as positive reinforcement and a reward upon completion. }
    \label{fig:AR-activity}
    \Description{}
\end{figure}

The AR activity embedded in the recognition phase (\autoref{fig:procedure}-d) aims to teach children about emotional expressions in an engaging and interactive manner (\autoref{fig:AR-activity}).
\name{} overlays a set of digital facial expressions of the chosen cartoon character onto the real-world environment that draws their attention effectively, and lets the child find a target emotion, creating an enjoyable way to allow children to learn emotional expressions in a familiar and safe-feeling background (\autoref{fig:UI-flow}-d).
We leveraged the ARKit XR Plugin and AR Foundation packages on Apple's iOS SDK that provide native AR integration with Unity's multi-platform XR API to implement this feature. 

The AR activity is a game structured into five rounds, with each round corresponding to an increasing level of difficulty, in terms of the number of expressions and the speed of movement they perform.
To add an element of challenge and maintain engagement, each round is limited to three minutes. 
At the start of the game, the system initializes by displaying four distinct facial expressions around a virtual sphere, centered from the child's perspective.
These expressions represent different emotions the child needs to learn to recognize.
The child then holds the tablet, using it as a dynamic window to peek into the surrounding world through the tablet's camera. 
The objective is to identify and ``capture'' the target emotion set in previous phases. 
To capture an emotion, the child must align the selected expression within a center area on the screen and hold it steady for over one second.
Upon capturing it, the system immediately analyzes the child's selection. 
If it is correct, \name{} provides positive feedback (\eg, cheerful animation of the expression and hearts) to enforce the learning. 
If the selection is incorrect, the system encourages the child to try again until timeout.

%% file: tables/cartoon-faces.tex
\begin{table*}[tb]
\caption{FACS-based \cite{Ekman1978FAC} emotional expression designs in \name{}.}
\centering
\vspace{-3mm}
\small
\label{tab:emotional-faces}

\begin{tabular}{lp{4cm}p{1.5cm}p{1.5cm}p{1.5cm}p{1.5cm}p{1.5cm}}
\toprule
\textbf{Emotion} & \textbf{FACS Description} & \multicolumn{5}{c}{\textbf{Cartoon Emotional Faces}}\\ 
\midrule
\textbf{Happiness} & Cheek Raiser, Lip Corner Puller 
& \raisebox{-.5\height}{\includegraphics[width=1.5cm, height=1.5cm]{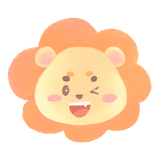}} 
&\raisebox{-.5\height}{\includegraphics[width=1.5cm, height=1.5cm]{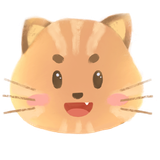}} 
&\raisebox{-.5\height}{\includegraphics[width=1.5cm, height=1.5cm]{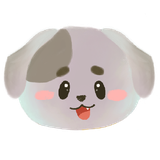}}
&\raisebox{-.5\height}{\includegraphics[width=1.5cm, height=1.5cm]{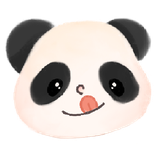}}
&\raisebox{-.5\height}{\includegraphics[width=1.5cm, height=1.5cm]{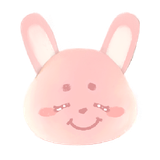}}\\

\textbf{Sadness} & Inner Brow Raiser, Brow Lowerer, Lip Corner Depressor 
& \raisebox{-.5\height}{\includegraphics[width=1.5cm, height=1.5cm]{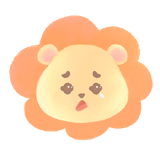}} &\raisebox{-.5\height}{\includegraphics[width=1.5cm, height=1.5cm]{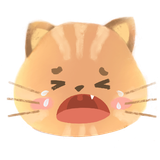}} 
&\raisebox{-.5\height}{\includegraphics[width=1.5cm, height=1.5cm]{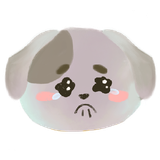}} 
&\raisebox{-.5\height}{\includegraphics[width=1.5cm, height=1.5cm]{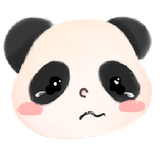}} 
&\raisebox{-.5\height}{\includegraphics[width=1.5cm, height=1.5cm]{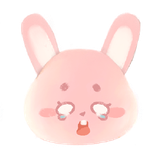}}\\

\textbf{Surprise} & Inner Brow Raiser, Outer Brow Raiser, Upper Lid Raiser, Jaw Drop 
&\raisebox{-.5\height}{\includegraphics[width=1.5cm, height=1.5cm]{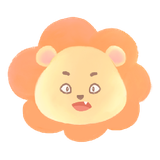}} 
&\raisebox{-.5\height}{\includegraphics[width=1.5cm, height=1.5cm]{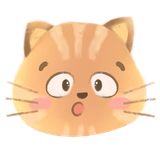}}   
&\raisebox{-.5\height}{\includegraphics[width=1.5cm, height=1.5cm]{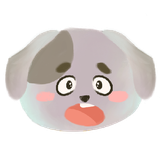}}
&\raisebox{-.5\height}{\includegraphics[width=1.5cm, height=1.5cm]{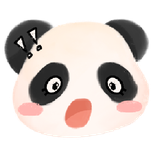}}
&\raisebox{-.5\height}{\includegraphics[width=1.5cm, height=1.5cm]{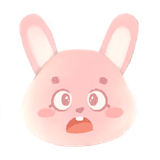}}\\

\textbf{Disgust} & Nose Wrinkler, Lip Corner Depressor, Lower Lip Depressor 
& \raisebox{-.5\height}{\includegraphics[width=1.5cm, height=1.5cm]{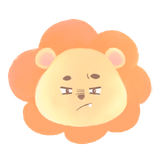}} 
&\raisebox{-.5\height}{\includegraphics[width=1.5cm, height=1.5cm]{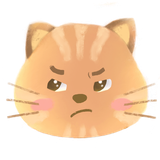}}  
&\raisebox{-.5\height}{\includegraphics[width=1.5cm, height=1.5cm]{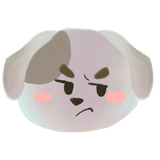}}
&\raisebox{-.5\height}{\includegraphics[width=1.5cm, height=1.5cm]{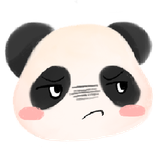}}
&\raisebox{-.5\height}{\includegraphics[width=1.5cm, height=1.5cm]{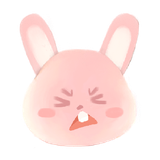}}\\

\textbf{Anger} & Brow Lowerer, Upper Lid Raiser, Lid Tightener, Lip Tightener 
& \raisebox{-.5\height}{\includegraphics[width=1.5cm, height=1.5cm]{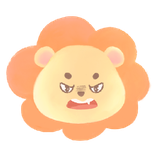}} 
&\raisebox{-.5\height}{\includegraphics[width=1.5cm, height=1.5cm]{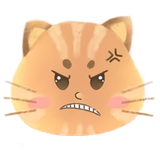}}   
&\raisebox{-.5\height}{\includegraphics[width=1.5cm, height=1.5cm]{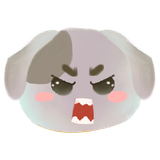}}
&\raisebox{-.5\height}{\includegraphics[width=1.5cm, height=1.5cm]{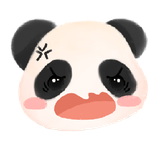}}
&\raisebox{-.5\height}{\includegraphics[width=1.5cm, height=1.5cm]{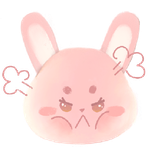}}\\

\textbf{Fear} & Inner Brow Raiser, Outer Brow Raiser, Brow Lowerer, Upper Lid Raiser, Lid Tightener, Lip Stretcher, Jaw Drop 
& \raisebox{-.5\height}{\includegraphics[width=1.5cm, height=1.5cm]{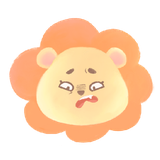}}  
&\raisebox{-.5\height}{\includegraphics[width=1.5cm, height=1.5cm]{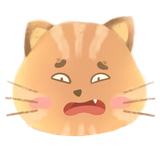}} 
&\raisebox{-.5\height}{\includegraphics[width=1.5cm, height=1.5cm]{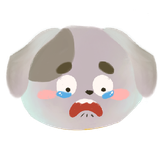}} 
&\raisebox{-.5\height}{\includegraphics[width=1.5cm, height=1.5cm]{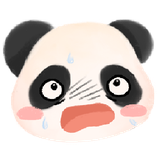}} 
&\raisebox{-.5\height}{\includegraphics[width=1.5cm, height=1.5cm]{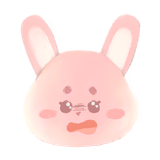}}\\

\textbf{Neutral} & In a natural state, facial expressions involve no muscle activity
&\raisebox{-.5\height}{\includegraphics[width=1.5cm, height=1.5cm]{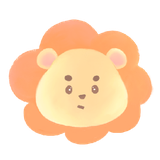}} 
&\raisebox{-.5\height}{\includegraphics[width=1.5cm, height=1.5cm]{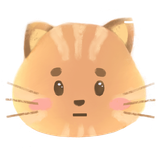}} 
&\raisebox{-.5\height}{\includegraphics[width=1.5cm, height=1.5cm]{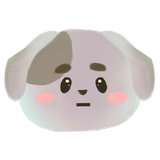}}
&\raisebox{-.5\height}{\includegraphics[width=1.5cm, height=1.5cm]{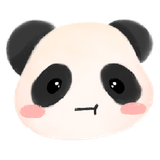}} 
&\raisebox{-.5\height}{\includegraphics[width=1.5cm, height=1.5cm]{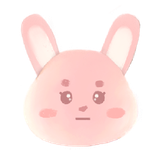}}\\
\bottomrule
\end{tabular}
\end{table*}

%% file: texfiles/6-evaluation.tex
\section{Evaluation} \label{sec:evaluation}
We conducted a between-subject controlled experiment \footnote{The study has received approval from the research ethics office at the authors' institution, ensuring compliance with ethical standards for research involving human subjects.}
to empirically examine how autistic children, alongside their caregivers, interact with \name{} within a real-world context, compared with a baseline. 

\name{} serves to probe innovative methods in augmenting social-emotional interventions through the application of AI and AR.

\subsection{Participants}
We recruited 24 autistic children (age: \mean{6.0}, \sd{2.3}; 3 girls and 21 boys) along with their caregivers from two special education centers.
The study was conducted in a safe environment, with secure data protection and debriefing sessions for caregivers. 
We aimed to resemble the environments and situations that children encounter daily during their normal sessions at the centers.
Informed consent was obtained from all caregivers or legal guardians on behalf of the children, given their age.
All recruited children possessed the cognitive capacity to engage in fundamental verbal communication with their caregivers and comprehend basic emotions.
These centers are recognized for offering specialized programs aimed at addressing the needs of autistic children and enhancing their social skills. 
Each child engaged in the study was accompanied by a caregiver, typically a staff member at the centers and occasionally a guardian.
The involvement of caregivers and guardians is essential, ensuring consistent collaboration and supervision in line with our system design for a supportive learning environment tailored to the needs of autistic children.
The study was structured to evaluate \name{} within a setting that replicates the typical conditions of daily educational sessions for ASD children.

\subsection{Baseline Condition}
We introduced a baseline that employed the social-emotional intervention using traditional teaching methods.
In particular, we curated slide decks that include the following components: storytelling, emotional face showcasing, emotion identification with facial expression cards, and question-and-answer exercises.
These traditional approaches do not incorporate the dynamic and immersive qualities offered by \name{}; however, the slide decks used the same cartoon characters and their emotional faces, as well as pre-generated stories from the same AI model. 
Moreover, we consulted with our experts and refined our slide decks based on their suggestions to ensure the baseline was of high quality and resembled their current teaching methods. 

\subsection{Procedure}  
\begin{figure}[tb!]
    \centering
    \includegraphics[width=0.32\linewidth,trim=0 10 0 10,clip]{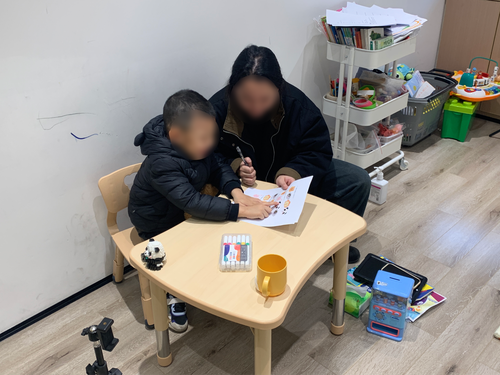}
    \includegraphics[width=0.32\linewidth,trim=0 10 0 10,clip]{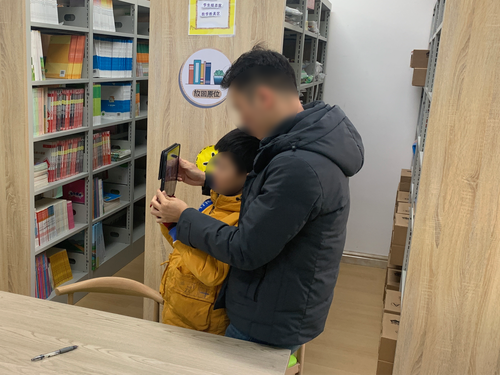}
    \includegraphics[width=0.32\linewidth,trim=0 10 0 10,clip]{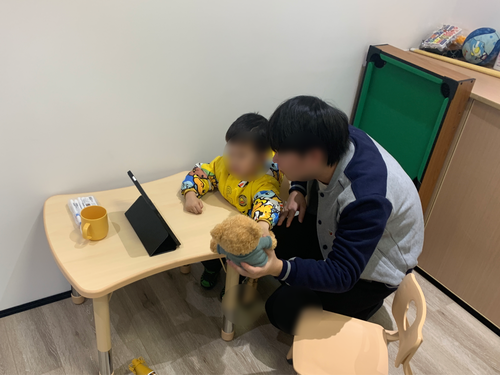}
    \vspace{-3mm}
    \caption{ Study sessions of \name{} with caregivers. Left: A child was pointing to the answer during a quiz. Center: A child and his father were playing together in the AR activity. Right: A child was listening to the story generated by the system and told by his father. %
    }
    \label{fig:onsite-pic}
    \Description{}
\end{figure}

The study was completed in a single session at the education centers for each child and caregiver pair.
Following the acquisition of consent from the caregivers, the study's objectives and methodology were explained to all participants. 
Then, the children were invited to take a pre-study quiz to assess their emotion recognition skills, consisting of 10 questions (\eg, Which one of the following expressions shows the little animal feeling disgusted?) randomly drawn from a prepared question bank.  
Next, the participants experienced the social-emotional intervention in a randomly assigned condition (\name{} or Baseline) and with an emotion of their choice for the intervention. 
For \name{}, each child was provided with an iPad with \name{} pre-loaded, and for Baseline, an iPad with the prepared slide decks was presented.  
For \name{}, we also prepared five objects in advance, ensuring a diverse range of items typically found at home and school. 
This selection included toys, cups, and classroom items.
By providing a familiar context, we aimed to make the gaming process intuitive and engaging for the children.
After the intervention, the children took a post-study quiz which also consisted of 10 questions from the same question bank.  
In the end, the caregivers completed a questionnaire; and for those in the \name{} condition, they also engaged in a semi-structured interview to gather qualitative feedback on their experience.
The whole study session lasted approximately 40 minutes. Each child received \$20 for their time and effort, and so did each caregiver receive \$30. %
The study sessions were video/audio-recorded for subsequent analysis.
\autoref{fig:onsite-pic} shows several study sessions with our participants on-site.

%% file: texfiles/7-results_quan_analysis.tex
\section{Results} \label{sec:results}
In this section, we discuss the findings from our evaluation, including both quantitative measurements and qualitative feedback.
In the following, we regard the children in the \name{} condition as E1-12 and those in the Baseline condition as B1-12, and the caregivers we interviewed in the \name{} condition as T1-12.

\subsection{Effectiveness in Emotion Recognition}

In our study, the children completed a pre- and post-quiz for the intervention they experienced (\ie, \name{} or Baseline). 
The quizzes focused on assessing their abilities in recognizing the emotion before and after the intervention. 
The score was out of 10, with each correct answer to a multi-choice question counting 1 point.
As shown in \autoref{tab:performance}, children with \name{} exhibited larger progress, with an average improvement of 1.5 points, contrasting with Baseline, with a mean decrease of 0.41 points. 
An unpaired t-test indicated that there was a significant difference in the change of scores between \name{} and Baseline ($t = 2.634$, $p = 0.015$).
This result highlights \name{}'s effectiveness in advancing the recognition and comprehension of emotions.
Without surprise, participants showed varying degrees of improvement across different emotions in the \name{} condition, with some children making significant strides, particularly in recognizing anger and surprise, and some children decreasing 1 point in sadness and disgust. 
This is plausible because of the short time period for our study; we believe \name{} could generate stabler positive improvement in the long run.
Additionally, score reduction was observed in the baseline condition.
Due to the limited number of participants, it is challenging to draw definitive conclusions. 
However, one explanation could be that children with the baseline might find the intervention's lack of engagement and interest, which might affect their attention and performance in the quiz afterward.
This emphasizes the importance of engagement in social-emotional interventions for autistic children.

\begin{table*}[tb]
\caption{Pre- and post-quiz scores for both \name{} and Baseline, along with the emotions of the interventions for participants.}
\centering
\vspace{-3mm}
\small
\label{tab:performance}
\begin{minipage}[b]{0.45\linewidth} %
\centering
\small
\begin{tabular}{lrrrl}
 \toprule
 \textbf{ID} & \textbf{Pre Score} & \textbf{Post Score}  & \textbf{$\Delta$Score}  & \textbf{Emotion} \\ 
 \midrule
E1 &  5 & 9 & \textcolor{green}{4}& happiness \\ 
E2 & 5  & 7 & \textcolor{green}{2} & anger \\
E3 & 3  & 9 & \textcolor{green}{6} & anger \\ 
E4 & 2  & 2 & \textcolor{yellow}{0} & neutral\\ 
E5 & 4  & 5 & \textcolor{green}{1} & fear \\  
E6 & 7  & 8 & \textcolor{green}{1} & fear \\ 
E7 & 7  & 9 & \textcolor{green}{2} & sadness\\  
E8 & 6  & 5 & \textcolor{red}{-1} & sadness \\ 
E9 & 5  & 4 & \textcolor{red}{-1} & disgust\\  
E10 & 3  & 5 & \textcolor{green}{2}& disgust\\  
E11 & 4  & 4 & \textcolor{yellow}{0} & surprise \\  
E12 & 7  & 9 & \textcolor{green}{2} & surprise \\   
\midrule
\textbf{M} & 4.83
& 6.33 & 1.50 & \\
\textbf{SD} & 1.70
& 2.46 & 2.02 & \\
\bottomrule
\end{tabular}
\subcaption{\name{}}
\label{tab:emostory-performance}
\end{minipage}%
\quad
\begin{minipage}[b]{0.45\linewidth} %
\centering
\small
\begin{tabular}{lrrrl}
 \toprule
 \textbf{ID} & \textbf{Pre Score} & \textbf{Post Score}  & \textbf{$\Delta$Score} & \textbf{Emotion}\\ 
 \midrule
B1 & 5  & 3 & \textcolor{red}{-2} & fear\\ 
B2 & 5  & 7 & \textcolor{green}{2} & happiness\\
B3 & 7  & 4 & \textcolor{red}{-3} & sadness\\ 
B4 & 3  & 3 & \textcolor{yellow}{0} & neutral\\ 
B5 & 3 & 1 & \textcolor{red}{-2} & anger\\  
B6 & 7  & 6 & \textcolor{red}{-1} & disgust\\ 
B7 & 5  & 7 & \textcolor{green}{2} & surprise\\  
B8 & 10  & 10 & \textcolor{yellow}{0}  & anger\\ 
B9 & 7  & 7 & \textcolor{yellow}{0} & sadness\\  
B10 & 4  & 3 & \textcolor{red}{-1} & surprise\\
B11 & 3  & 3 & \textcolor{yellow}{0} & disgust\\
B12 & 6  & 6 & \textcolor{yellow}{0} & neutral\\
\midrule
\textbf{M} & 5.41
& 5.00 & -0.41 & \\
\textbf{SD} &2.11 
& 2.56 & 1.51 & \\
\bottomrule
\end{tabular}
\subcaption{Baseline}
\label{tab:baseline-performance}
\end{minipage}%

\end{table*}

\subsection{Questionnaire Responses} 
\begin{figure}
    \centering
    \includegraphics[width=1\linewidth]{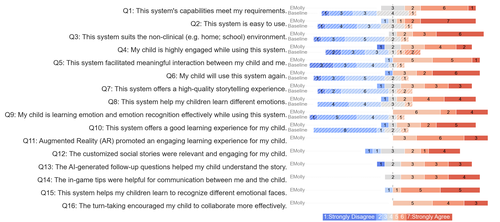}
    \vspace{-8mm}
    \caption{Participants' ratings on the questionnaire, where Q1-3 regard usability, Q5-10 regard general experience, and Q11-16 regard the experience of proposed features in \name{} (1: Strongly disagree, 7: Strongly agree).}
    \label{fig:likert}
    \Description{}
\end{figure}

The questionnaire was structured to assess the usability and user experience for both conditions.
Additionally, for \name{}, we collected users' reception of the proposed features including customization, interactivity, and impact of integration AI and AR. 
Responses were recorded on a 7-point Likert scale as shown in \autoref{fig:likert}.
Medians of the ratings for each question were computed and Mann-Whitney U tests were performed to compare the two conditions. The results indicate that \name{} significantly outperformed the baseline on all the comparable aspects (Q1-Q10).

Higher ratings were observed across different usability aspects of \name{}, compared to Baseline; participants thought \name{} was better in meeting the requirements (Q1: $Mdn_E=6$, $Mdn_B=3$, $U=134.0$, $p < 0.001$), easier to use (Q2: $Mdn_E=7$, $Mdn_B=3$, $U=135.5$, $p <0.001$), and more ubiquitous (Q3: $Mdn_E=6.5$, $Mdn_B=3$, $U=141.0$, $p <0.001$).
On the user experience, the level of engagement of \name{} 
\rev{trended higher than that of Baseline, but these differences were not statistically significant (Q4: $Mdn_E=5.5$, $Mdn_B=3$, $U=114.5$, $p=0.014$).}
There was one participant (E8) who rated it lower, potentially attributed to a lack of familiarity with the game's theme and thus not engaging with the stories.
Meaningful interactions were better facilitated between caregiver and child by \name{} (Q5: $Mdn_E=5.5$, $Mdn_B=2$, $U=137.5$, $p <0.001$), demonstrating that \name{} can transcend educational utility and foster deeper connections during the learning process. 
This enhanced engagement was also mirrored in the expressed willingness of participants to return to the system (Q6: $Mdn_E=6.5$, $Mdn_B=3$, $U=133.0$, $p <0.001$). %
\name{}’s storytelling experience was regarded as superior to Baseline (Q7: $Mdn_E=5.5$, $Mdn_B=3$, $U=126.0$, $p =0.002$), offering a more compelling narrative experience\rev{, but these differences were not statistically significant}.
It also excelled in enabling children to learn and distinguish different emotions (Q8: $Mdn_E=6$, $Mdn_B=3$, $U=130.5$, $p < 0.001$) and facilitated more effective teaching of emotional recognition (Q9: $Mdn_E=5.5$, $Mdn_B=3$, $U=141.0$, $p <0.001$) than the Baseline condition.
Furthermore, the learning experience provided by \name{} (Q10: $Mdn_E=6$, $Mdn_B=2$, $U=132.5$, $p <0.001$) was considered superior. 

For the proposed features in \name{}, AR was perceived to be effective in enhancing the engagement (Q11: $Mdn_E=6$), and the customized social stories were well-received (Q12: $Mdn_E=5$), affirming the value of allowing children to personalize their experience by combining the surroundings into the narratives in an immersive setting.
Yet, two participants suggested potential areas for refinement in the customization process to have a deeper reflection of diverse preferences within autistic children.
For example, T5 suggested incorporating familiar elements, such as family members, into the game to enrich the personalization further.
The results also show that the AI-generated follow-up questions enhanced children's understanding of emotions (Q13: $Mdn_E=5.5$), and the communication tips provided valuable guidance for caregiver-child interactions (Q14: $Mdn_E=6$). 
Moreover, \name{}'s support in teaching the recognition of various emotional faces was perceived as highly effective (Q15: $Mdn_E=6$).
The turn-taking mechanism, encouraged by \name{}, was particularly noted for improving collaborative efforts (Q16: $Mdn_E=6$).

%% file: texfiles/8-results_qual_analysis.tex
\subsection{Interview Results}
A comprehensive thematic analysis was conducted on the qualitative data obtained from semi-structured interviews with the 12 caregivers. 
All audio recordings of interviews were transcribed verbatim. 
These transcriptions, along with the audio recordings, we conducted open coding to identify initial themes, followed by emergent coding to link themes and identify patterns. 
We then grouped the initial themes to distill the main insights, iteratively reviewing and refining the codes to ensure consistency and reliability.
The goal was to establish systematic understandings of \name{} features guided by the design principles, focusing on aspects such as caregiver-child collaboration, customization, and how AI and AR impact user experience and learning outcomes.
In this section, we discuss our findings with the following themes, in light of the previous quantitative results.

\textbf{\name{} fosters caregiver-child collaboration (D1).}
Overall, the feedback underscores the critical role of caregiver-child collaboration in ASD social-emotional learning interventions, with \name{} designed to nurture participation, communication, and cooperative interaction among children and their caregivers.
T1 appreciated the system's user-friendly design for home setting, noting its ease of use and the dynamic, interactive appeal that could capture children's interest and encourage more profound participation from both children and caregivers.
T10 applauded \name{}'s interactive prompts that encouraged children's active participation, making the learning process both rich and enjoyable without needing her constant explanations.
Moreover, the dynamic nature of AI-generated stories was praised for its convenience, adaptability, and richness, offering caregivers a valuable resource in engaging children with diverse, captivating narratives without the heavy lifting of creating content themselves (T12). 
This was echoed by T1, who highlighted \name{}'s effectiveness in fostering conscious thought about collaboration and turn-taking without overt effort.
This sentiment along with the importance of verbal communication in the learning process as emphasized by T6: \name{} \pqt{fostered a shared space for communication through storytelling}{T6}.
Further, it was agreed that \name{} \pqt{offered opportunities for collaborative interaction within an engaging game format, which is something he (E3) was happy to participate in together with me,}{T3} and \pqt{during the Q\&A, although it was somewhat mandatory for him to respond, it created opportunities for interaction. }{T12} emphasizing \name{}'s ability to facilitate caregiver-child communication and interaction.

Moreover, the caregiver involvement provided by \name{} was appreciated, with T1 and T2 highlighting how the system's interactive experiences offered opportunities for meaningful collaboration. 
\pqt{This collaborative aspect was not only crucial for facilitating emotional learning but also seemed to strengthen familial bonds as children and caregivers navigate the system together.}{T1} 
T2 also mentioned the act of making expressions together helped children better understand emotions, with caregivers playing a pivotal role in demonstrating and guiding these expressions in a collaborative method.
Furthermore, T4 underscored the significance of real-life interactions over static imagery by pointing out, \qt{It's ineffective to only see happy faces on cards... You need to witness emotions through in-person communication, observing happiness directly.}
This comment reveals the critical importance of direct, experiential learning and interaction in emotional recognition and understanding.

Participants also highlighted the thoughtful design and adaptability of \name{} in AI generation, allowing them to effortlessly produce suitable stories for the intervention. 
T12 noted the importance of setting task difficulty appropriately, emphasizing it ensures a more tailored and efficient learning experience. 
T1 appreciated the customization aspect and suggested that the story difficulty could start at different levels for children with different proficiency, which could be a future feature to implement to better adapt to the children's learning pace. 
These comments underscore \name{}'s capacity to provide engaging AI-generated stories while making the process effortless for caregivers.

\textbf{\name{} promotes and sustains engagement through interactivity (D2).}
\name{} is equipped with the ability to keep children's interest in social-emotional learning through a blend of AI-enabled interactivity, dynamic social stories and immersive AR.
This aspect is vital for fostering a long-term commitment to social-emotional intervention for children with ASD. 

T1 emphasized the dynamic nature of these games as more attractive and involving for both children and parents, stating, \qt{Interactive games are definitely dynamic...more appealing, making children more interested, or even parents more involved.} 
Moreover, \name{} offered children the opportunity to try making expressions themselves, which was more effective than the teaching approach offered by books or cards and providing AI-enabled real-time responses.
The participants also highlighted the AI's capacity to create an interactive learning environment, enhancing interactivity and deepening emotional comprehension. 
As T2 noted, \qt{He enjoys the AI's emotional feedback}, emphasizing the value of real-time responses in engaging children.
This real-time feedback coupled with the dynamic and diverse social stories empowered by the AI were praised by T3: \qt{creating an interactive learning environment, enhanced interactivity and contributed significantly to a deeper understanding of emotions.}  
 
The immersive and interactive nature of AR in \name{} continued to be a focal point, particularly for its role in enhancing engagement and learning outcomes by providing a realistic and immediate experience. 
The use of AR to create novel learning experiences was preferred over traditional methods such as cards and books, garnering positive feedback from both children and caregivers. 
The entertainment and immersion offered by AR games were noted to make the teaching of emotional recognition more captivating and interactive. 
T12 shared, \qt{AR is interesting and immersive, enhancing children's learning participation, which helps with learning outcomes.} 
Further, the immersive experiences facilitated by AR were recognized for increasing children's engagement in learning, thereby supporting educational achievements.
Incorporating additional insights, children's delight in the game’s playfulness was evident. 
T7's enthusiasm, \qt{I want to play this game again with big sister. Why? Because it's so fun to play games. I like finding expressions the most; they're interesting with so many expressions,} revealed the joy and curiosity AR stimulates. 
T1 observed the potential for achievement and comprehension through repetition, \qt{He's quite interested in finding emotional expressions. With more practice, he might gain a better understanding and a sense of accomplishment.} 
T12 reflected on the blend of reality and gameplay AR offered, \qt{It feels like he's playing within the app, but AR allows him to search in the real world... This reality-based interaction could enhance the sensory experience.} 
T12 also appreciated how AR connected with the real world, offering a novel experience that diverges from traditional learning methods. 
T6 noted the higher level of investment children have when playing augmented reality games, emphasizing the immersive aspect of participation.

Comparatively, T1 discussed AR’s advantages and potential limitations against traditional books, stating, \qt{The advantage definitely lies in its interactivity and fun, though it requires digital devices and setup. Books might be more straightforward to use but offer a simpler recognition process. Games, in contrast, allow for a multifaceted approach, making the learning outcome more effective.}
T4 mentioned the quicker pace of AR games, which may sacrifice detail for speed, yet the high engagement and enjoyable nature make it a preferable option for children, \qt{The fast pace might miss finer details, but the fun and interactive game format keeps children more willing to participate.}
T4 also highlighted the focused attention children give to AR games, \qt{Kids pay more attention when playing games. They're not just playing; they're learning by looking for specific expressions, which is an effective way to learn through play.}

\textbf{\name{} personalized learning experiences (D3).}
By aligning game elements with individual children's interests and developmental stages, \name{} can offer a dynamic, optimized learning experience.
T12 highlighted the unique appeal of the customization, favoring the \name{}'s ability to generate diverse social stories based on objects chosen by the children themselves. 
\pqt{It's quite interesting... it generates different social stories based on the scanned objects, allowing the child to guide the story.}{T12} 
It indicates that this approach not only captures the children's interest but also empowers them to shape their learning materials.
T2 also underscored the innovative aspect of creating unique social stories from everyday items, saying, \qt{If he's interested in an item, and there's a story involving that item, he's more willing to listen to the story. It's like how children search on apps for what they like.}
Further enhancing engagement, T12 observed the meaningful impact of choosing items to be included in stories: \qt{When he selects an item, and it becomes part of a story, it's meaningful. He can know what story happens with that item, which is quite good and connects well.}
It was agreed that this process not only appeals to the children's curiosity but also deepens their sense of involvement and participation in the learning experience.
T5 reflected on the simplicity and appropriateness of the stories, \qt{It’s actually quite straightforward... quite simple, suitable for young children, and appropriately challenging. It also incorporates knowledge beyond emotions, which is quite fitting.}

This adaptability starkly contrasts with more rigid formats, such as storybooks, which T12 noted that it could not be altered on the fly, emphasizing that \qt{Flexibility is definitely better compared to fixed formats.} 
The dynamic nature of \name{}'s storytelling could allow for a richer narrative experience. 
T6 found value in this feature, stating, \qt{It's good because you can enrich your story... Many things might initially capture their interest, but they may lose interest upon further exposure. Being able to switch stories if they don't like one is beneficial.} 
T5 also highlighted the distinctive advantage of \name{} over other story apps, noting its variability: \qt{The stories are changeable, not monotonous, which is different from other apps.}

Insights from T6 and T5 emphasized the importance of considering the unique and sometimes unpredictable interests of children with ASD. 
\pqt{You have to consider some things you might not think of because their interests can be very specific... If you can incorporate these unique interests into social stories, they (children) will probably like them more.}{T6}
T5 added the possibility of including more personalized elements, \qt{If facial expressions could also be customized, it would be more flexible (for learning).}
This feedback underscores the significance of customization in enhancing the learning experience for children with ASD.

%% file: texfiles/9-discussion.tex
\section{Discussion}\label{sec:disscussion}

To inspire broader explorations on designing social-emotional intervention to promote penalization, caregiver involvement, and engagement in the future, we formulate a set of design implications as follows.

\textbf{Implication 1: Easing the integration of personally relevant elements into customized interventions.}
Our system effectively incorporates objects of personal significance to the participating autistic children, 
into customized social stories, which were valued by the caregivers and shown to be a meaningful approach to engaging autistic children (D3).
Expanding beyond ASD support, recent research has begun exploring AI-driven personalization in broader contexts with neurotypical children. 
Liu et al. and Han et al. used AI to enhance children's emotional and creative expression, incorporating personal toys into family stories and employing AI-powered visual storytelling, respectively~\cite{liu2024he, han2023design}.
Given the proven benefits for both neurotypical and ASD children, future research should explore integrating personal elements into stories and gameplay.
Advancements could simplify capturing and uploading images or 3D scans of familiar objects, with image-based GenAI and LLMs transforming these into story elements.

\textbf{Implication 2: Supporting more adaptive design for social-emotional learning for individuals with ASD.} 
Given the varying functioning levels of young autistic children, a flexible and personalized intervention design is essential~\cite{Lyu2023}. 
We implemented a system that generates social stories with varying emotional and scenario complexity to meet diverse needs (D3). 
Caregivers have particularly appreciated the simplicity and appropriateness of these stories for young children, enriching the intervention experience.
We recommend designers collaborate with experts to fine-tune social story difficulty levels for better user adaptiveness. 
Integrating a personalized AI assistant, as explored in~\cite{xi2023rise}, could enhance adaptability through accumulated child-AI interactions. 
This technology could record and analyze interactions in real time, tailoring social stories to the verbal and cognitive functioning levels of each child.

\textbf{Implication 3: Considering child-caregiver turn-taking and collaborative play in ASD interventions.} 
It is crucial to involve caregivers in interventions for autistic children, as it significantly impacts long-term progress~\cite{Chaidi_Drigas_2020}. 
In our gameplay, caregivers performed facial expressions linked to the stories, serving as models for the children (D1). 
Our findings highlight \name{}'s effectiveness in facilitating collaborative learning of facial expressions through turn-taking.
We recommend future designs incorporate more multimodal turn-taking mechanisms to engage both caregivers and autistic children. 
For example, integrating visual cues with auditory feedback, as discussed in~\cite{huang2022starrescue}, could enhance the effectiveness of social-emotional interventions.

\textbf{Implication 4: Promoting children's engagement in social-emotional learning through interactivity and feedback.}
In our study, to motivate children to learn emotions and promote engagement, we developed an interactive AR activity that allows children to interact with digital facial expressions in the real world (D2). 
This blend of realistic and virtual experiences enhances sensory engagement and learning.
Inspired by our findings and prior research on AR games for nonspeaking autistic people~\cite{nazari2023interactive}, we suggest designers include more nonverbal, embodied animations, such as jumping and somersaults. 
Additionally, incorporating fidgeting behaviors like swiping, tapping, and clicking with supportive visual and auditory feedback~\cite{ji2022ar} could help young autistic children perform soothing behaviors and enrich their motor-sensory experiences.

\textbf{Limitations and Future Work.} 
\name{} has demonstrated effectiveness, and our results provide rich implications for future practice. However, there are several limitations in our study.
One crucial consideration is the safe use of large language models (LLMs) with social stories for young autistic children. 
We selected GPT-4 for its filtering mechanisms, but unintended biases or misunderstandings may still arise, especially given the unique needs of autistic children. 
Future research should incorporate more professional literature on social stories and ASD intervention into LLMs and collaborate closely with parents, experts, autistic children, and stakeholders to ensure the safe utilization of GenAI.
Another limitation is our modest sample size. 
Although consistent with prior research, individual differences among participants with ASD mean our findings may not apply universally~\cite{porayska2018blending}. Expanding to larger field studies is essential for a comprehensive understanding of social-emotional interventions in ASD~\cite{Lyu2023}. Future research should conduct long-term studies with diverse participants, considering various functioning levels and cultural contexts.
Lastly, our game characters were all cartoon animals, which, while attractive, may not be as effective as familiar characters. Caregivers recommended using characters representing familiar people or objects to strengthen connections between the game and reality. Future designs should investigate and compare different representations and offer customization options for caregivers and experts, allowing them to tailor game elements and scenarios to meet specific needs.

%% file: texfiles/10-conclusion.tex
\section{Conclusion}
We have introduced \name{}, an interactive tablet game that enhances the social-emotional intervention of autistic children and actively involves caregivers.
\name{} seamlessly integrates AI-generated content with AR activities, providing a dynamic and immersive learning experience that captivates children's interests and supports their social-emotional development.
\name{} was developed over an iterative co-design process involving domain experts, caregivers, and children. %
Based on the co-design process and prior art, we derived a number of insights and established a set of principles for the design of social-emotional intervention tools for autistic children. 
Compared with a baseline, we assessed \name{} through a between-subject controlled study with 24 autistic children at two special education centers, accompanied by their caregivers. 
Quantitative and qualitative results indicate the effectiveness of \name{} and offer deep insights into the design and development of mobile AR and AI games as effective tools for social-emotional intervention in non-clinical settings.

%% file: texfiles/Appendix-A_StoryExamples.tex
 \section{Social Story Examples}\label{appendix:stories-examples}

 \begin{itemize}
    
    \item Lan and the Happy Painting Time with Colorful Crayons
    \begin{itemize}
        \item Character: Lan
        \item Emotion: Happiness
        \item Difficulty: Medium
        \item Story: While I was in the kindergarten painting corner, I found a box of colorful crayons. I wanted to draw a big sun and invited my friend Ming to draw with me. We took turns using different colors. Ming drew a yellow sun and I drew a blue sky. We both felt happy because working together made the painting more beautiful.
     \end{itemize}

    \item Jie's Jump Rope Contest
    \begin{itemize}
        \item Character: Jie
        \item Emotion: Happiness
        \item Difficulty: Medium
        \item Story: Jie won first place in the school's jump rope contest and felt very happy. His friends danced and cheered around him, celebrating his success. Jie learned that hard work and practice can lead to success, and sharing it with friends makes the victory even sweeter.
     \end{itemize}
 
    \item The Little Cat and the Damaged Toy Bear
    \begin{itemize}
        \item Character: Little Cat
        \item Emotion: Anger
        \item Difficulty: High
        \item Story: Little Cat returned home and found that her favorite toy bear was broken. She started feeling angry, thinking that her pet dog had damaged it. Little Cat took a deep breath and tried to stay calm. She decided to first check the toy bear and found that it was just the seams that had come loose. Her mom helped her fix the toy bear, and Little Cat learned to understand the whole situation before getting angry.
     \end{itemize}
     
    \item Li's Football Game
    \begin{itemize}
        \item Character: Li
        \item Emotion: Anger
        \item Difficulty: High
        \item Story: During a football game at school, Li got angry when a friend accidentally kicked his leg instead of the ball. His teacher noticed and taught him how to express displeasure without getting angry. Li told his friend how he felt angry, and the friend apologized. They shook hands and continued the game, and Li learned how to manage his anger.
     \end{itemize}

    \item Little Rabbit and the Rotten Apple
    \begin{itemize}
        \item Character: Little Rabbit 
        \item Emotion: Disgust
        \item Difficulty: Medium
        \item Story: During lunch at school, little Rabbit took an apple from her lunch box. After biting into it, she found a worm inside, which disgusted her. Her teacher noticed and helped her deal with the rotten apple and taught her the importance of checking the quality of fruits.  Little Rabbit learned how to choose fresh fruits to avoid similar situations.

     \end{itemize}

    \item Little Frog and the Mud Challenge 
    \begin{itemize}
        \item Character: Little Frog
        \item Emotion: Disgust
        \item Difficulty: High
        \item Story: Little Frog always felt disgusted by the mud in the pond. One day, his ball fell into the mud. Initially, he didn't want to retrieve it, but then he mustered up the courage, jumped into the mud, and got his ball back. Although the mud was dirty, Little Frog realized that sometimes it's necessary to face things we dislike. He learned to bravely confront his feelings of disgust and felt proud of his actions.
     \end{itemize}

    \item Bai's Drawing Time
    \begin{itemize}
        \item Character: Bai
        \item Emotion: Neutral / calm
        \item Difficulty: Low
        \item Story: During a weekend afternoon, Bai took out his colored pencils and began to draw on the balcony at home. He drew a beautiful garden scene, feeling very calm and content. His drawing attracted his family's admiration, which made him even enjoyed. This drawing experience made Bai decide to practice regularly to improve his drawing skills.
     \end{itemize}

    \item Hua's Reading Time
    \begin{itemize}
        \item Character: Hua
        \item Emotion: Neutral / calm
        \item Difficulty: Low
        \item Story: Hua found an interesting picture book in the library. He sat quietly in a corner and read intently. He felt calm and relaxed and enjoyed the time of reading. After finishing, Hua felt very good and decided to visit the library every week to read more books.
        
     \end{itemize}

   \item  Lu's Nighttime Adventure
    \begin{itemize}
        \item Character: Lu
        \item Emotion: Fear
        \item Difficulty: Low
        \item Story: Lu was afraid of the dark at night. When his room got very dark, he felt uneasy. His mother gave him a night light, which made him feel reassured. He turned on the night light, and the light in the room made him feel safe. Lu learned to overcome his fear with the help of the night light.
     \end{itemize}
     
    \item Lan's First Day of School
    \begin{itemize}
        \item Character: Lan
        \item Emotion: Fear
        \item Difficulty: Medium
        \item Story: Lan was a bit scared on his first day at school because everything was unfamiliar. He carried his new backpack filled with his favorite stationery and books. With the teacher's guidance, he gradually felt more at ease. He met some new friends and they explored the school together. By the end of the day, Lan was no longer scared but looked forward to the next day.
     \end{itemize}

    \item Zi's Jump Rope Surprise
    \begin{itemize}
        \item Character: Zi
        \item Emotion: Surprised
        \item Difficulty: Medium
        \item Story: During PE class at school, Zi tried single rope jumping for the first time. She initially thought she couldn’t do it, but with her friends' encouragement, she gave it a try. To her surprise, she managed to jump ten times and could keep going. This made her feel very excited and proud. Zi learned that trying new things can bring surprises and that a sense of achievement comes from not giving up.
     \end{itemize}
     
    \item Hong's Surprise Gift
    \begin{itemize}
        \item Character: Hong
        \item Emotion: Surprised
        \item Difficulty: Low
        \item Story: At the end of a school day, Hong received an anonymous gift box which surprised her greatly. Upon opening it, she found a book she had always wanted. Her friends gathered around and celebrated with her. Hong learned that sharing happy moments can enhance the joy of friendships.
     \end{itemize}

    \item Yu's Origami Contest
    \begin{itemize}
        \item Character: Yu
        \item Emotion: Sadness
        \item Difficulty: Medium
        \item Story: Yu participated in the school's origami contest. Despite his hard preparation, he did not win, which made him sad. His teacher and friends comforted him and told him that what matters is participation and trying new things. Yu understood that failure is part of learning and decided to keep practicing origami.
     \end{itemize}
     
    \item Hu and the Lost Colored Pen
    \begin{itemize}
        \item Character: Hu
        \item Emotion: Sadness
        \item Difficulty: High
        \item Story: Hu was very sad when she discovered her favorite blue colored pen was missing during art class. She began to search everywhere, but couldn't find it. Seeing her sadness, her friend Cheng decided to help her search. Finally, they found the pen in a corner of the classroom. Hu learned that having friends' support is important when facing difficulties.
     \end{itemize}

 \end{itemize}

%% file: texfiles/Appendix-B_ExpertsQuotations.tex
 \section{Experts Quotations}\label{appendix:experts-quotations}
 \begin{itemize}
    \item R1 Adaptive Story Complexity
    \begin{itemize}
        \item Stories need to be aligned with their (autistic children's) cognitive level and understanding. - E3
        \item In general, younger children who are just starting to learn about emotions may focus on the four basic emotions: happiness, anger, sadness, and fear. For these basic emotions, the follow-up questions of the story need to be simpler. - E5
        \item Tailor the stories to align with different stages of physical and mental development (of autistic children), ensuring that the stories are adaptable and extendable. - E4
    \end{itemize}
    
    \item R2 Familiar Story Scenarios
    \begin{itemize}
        \item By embedding familiar scenarios from daily life, children can more easily understand complex emotions. - E3
        \item Some children may not have seen or played with certain toys like LEGO. It might be better to consider toys that are more familiar and easily accessible to the children. -E1
    \end{itemize}
    
    \item R3 Diverse Story Elements
    \begin{itemize}
        \item When I gave (autistic) children a specialized lesson on emotions, I would prepare some cards. For example, I would use cards to teach emotions like disgust. - E5
        \item  Different emotions cannot be triggered in the same way. Emotions and facial expressions are different. - E1
        \item Consider the choice of toys or preferred items that have indeed been researched by some people. - E1
    \end{itemize}
    
    \item R4 Attractive Visual Illustrations
    \begin{itemize}
        \item (Autistic) Children really need visual support, such as the support of pictures. - E2
        \item The visualization in the process of (the story) is relatively important, he (child) will be able to understand the story better. - E3
        \item The key to visualization is that it needs to be dynamic. The changes and animations are likely more important for children's understanding of the story. - E1
    \end{itemize}
    
    \item R5 Emotional Reasoning and Resolution
    \begin{itemize}
        \item For children with higher level ability, the goal of teaching emotions is not just to recognize the emotion but also to understand the entire context in which the emotion arises, including the cause, the resulting outcome, and the behavior that the emotion might lead to. They need to understand others' emotions and then transfer this understanding to themselves. -E3
        \item In terms of scenarios, you (story) can provide more contexts. For example, if something causes disgust, will other objects or situations evoke the same emotion? - E2
        \item  They understand emotions based on outcomes. For example, if the mother says something and it leads to a bad outcome, it will accumulate in their mind. If she buys me candy, but then in the next moment, criticizes me, they link the two together and think, I can't take this because of what Mom said. - E2
    \end{itemize}
    
    \item R6 Interactive and Immersive Learning
    \begin{itemize}
        \item You can also set up different scenarios to simulate various situations, helping them deepen their understanding of emotions. This can include different people, environments, and types of games. Additionally, introducing different instructors or strangers to teach them about these emotions can be beneficial.- E4
        \item If we use augmented reality to highlight human facial expressions, it makes it easier for them to notice. By addressing their attention issues, they will be able to learn more effectively and achieve better learning outcomes. - E5
         \item In addition to observing facial expressions, it is also necessary to combine actual situational information to determine what happened. For example, in the classroom, understanding what event occurred and what emotions it triggered helps in comprehending the situation. - E4
    \end{itemize}
    
    \item R7 Multimodal Gamified Feedback
    \begin{itemize}
        \item Stories always have a process and can change across several pages. Sometimes there will be sound, like a rabbit's surprise or a teddy bear that makes a rumbling sound when you touch its head. This kind of thing can also include sound effects.- E3
        \item Apart from paying attention to this, it would be more effective if the facial expression was dynamic, moving and animated. -E4
    \end{itemize}

   \item R8 Familiar Learning Setup
    \begin{itemize}
        \item In a sense, it tends to be easier, and it (recognizing facial expressions) seems to be relatively easier with familiar people.- E1
        \item If it's a parent's or teacher's face, the child is more likely to pay attention. If it's a stranger's face, it's harder for them to judge the emotion or pay attention. They tend not to engage with a stranger. - E2
        \item If the environments (in the story) he (child) frequently encounters, such as the classroom, home, and places like restaurants, playgrounds, and amusement parks, he will come into contact with various settings. The distractions in these environments can affect him (with engagement). -E5
    \end{itemize}

    \item R9 Clear Learning Focus
    \begin{itemize}
        \item The factors that trigger emotions should be as simple and direct as possible (primitive reflex), and more universal. For example, physiological factors like feeling uncomfortable when hungry or being unhappy when someone takes away their food. - E1
         \item But sometimes for children with lower comprehension levels, they don't understand complex questions. For example, if you want him to find an angry expression, you can't ask, 'Where is the angry expression?' Instead, you might need to directly tell him, 'Point to the angry face.' Using the simplest words to prompt him might be more effective. - E2
    \end{itemize}
    
    \item R10 Pretend Play in Emotional Context
    \begin{itemize}
        \item Through contextual pretend play, children can better understand emotions. -E2
        \item We can set up different scenarios and stages during the teaching process to pretend play to help them learn about emotions, allowing them to continuously deepen their understanding of emotions. This can involve different people, environments, and types of games. -E3
    \end{itemize} 
\end{itemize}

%% file: texfiles/Appendix-C_Prompt.tex
\section{Story Generation Prompts}
The following prompts are designed to generate a social story. The main prompt focuses on creating an effective social story, while the additional consideration prompt allows for incorporating specific requirements, and the adjustment prompt is used to modify the difficulty level as needed.
\subsection{Main Prompt} \label{appendix:prompt}
Social stories are written from the perspective of children with autism, providing them with accurate social information. They can be used to describe and anticipate specific social situations, explain and understand social procedures and conventions, guide children through the steps and techniques of social activities, and prompt appropriate social responses. A social story includes the following types of sentences:
A. Descriptive Sentences: These sentences appear at the beginning of the social story. They describe the situation and the people involved, what is going to happen, and the reasons for the events. They also answer the questions: Where? Who? What will happen?
B. Perspective Sentences: These sentences describe internal feelings—how people feel, want, think, and believe in the situation. These sentences are crucial because the information they contain is inaccessible to children with autism spectrum disorders.
C. Directive Sentences: These sentences present social cues in the situation and indicate the expected response. Such responses may start with "I will try" or "I will attempt."
D. Control Sentences: These sentences are added to the story by the storyteller, describing general observations and thoughts to reinforce the information presented in the story.
E. Affirmative Sentences: These sentences emphasize the importance of directive sentences; they start with "It is good to..."
F. Cooperative Sentences: These sentences describe the actions of others, showing who they can help and how they can help.
When writing social stories, some elements must be included, and some principles must be followed:
Use basic sentence types such as descriptive, perspective, directive, and affirmative sentences, or add cooperative or control sentences as needed based on the case.
The most important rule is that the sentences in the social story should be in a certain proportion: two to five descriptive or perspective sentences plus no more than one directive sentence.
Please generate a social story with two to five sentences that are close to reality. 
The social scenario must be close to the daily life of children with autism, suitable for daily social behavior training.

The story must meet the following requirements:
Have a cartoon character as the protagonist, a common toy or item for children, and an emotion.

Follow the following rules when writing the story:
Describe the scene and characters: The social story should describe a specific scene and the people involved. For example,``When I play in kindergarten, I meet a new friend named Xiaoming''
Describe the situation and goal: Next, the social story should describe the current situation and goal. For example, ``I want to be good friends with Xiaoming''
Describe appropriate and inappropriate behavior: Next, the social story should describe appropriate and inappropriate behavior in the situation. For example, ``When playing with Xiaoming, I should smile, talk, share toys, not push him or grab his toys''
Describe feelings and consequences: The social story should also describe the feelings and consequences of different behaviors. For example, ``If I smile, talk, and share toys, Xiaoming will feel happy and play with me. If I push him or grab his toys, Xiaoming will feel unhappy and may not want to play with me again.''
Conclusion and positive behavior: Finally, the social story should summarize appropriate and inappropriate behavior and emphasize which positive behaviors should be taken. For example, ``When I want to be friends with Xiaoming, I should smile, talk, and share toys. In this way, Xiaoming will feel happy, and we can play together and become good friends.''

Note that there should be a social scenario in the story, and it should include a solution, a cartoon character as the protagonist, a common toy or item for children, and an emotion. 
The story can only include one of the following emotions: ``angry'', ``happy'', ``sad'', ``disgusted'', ``scared'', ``surprised'', or ``calm''. 
The story should be suitable for children with autism, about 30 words, and each sentence should not be too long than 20 words. 
Please give the story a difficulty level, either low, medium, or high, considering the used language, vocabulary, sentences, social scenarios, interactions that in the stories. 
The low difficulty story is a simple, straightforward story with basic vocabulary and clear, predictable plot developments with short sentences. Focus on a single character facing a simple problem, like sharing toys, with a direct solution. Maintain consistent attitudes and preferences is advised to avoid confusing children with different emotional responses.
Medium difficulty is a story with slightly more complex language and a bit more nuanced plot. Include characters and introduce basic emotional expressions and reactions, such as resolving a misunderstanding between friends. 
High difficulty story is a story with multiple plot twists, richer vocabulary, and advanced emotional expressions. Involve several characters interacting in a more complex social situation, such as collaborating on a group project, which requires understanding and applying more social skills. 

After the story, please generate five questions that can be used to communicate with the child and their answers to review the emotional social scenarios and solutions in the story, with the recap question related to the child's own emotions in the story. 
The results are output according to the following format:{"role": "user", "content": "Emotion: Character: Item: Story Difficulty: "}, 
{"role": "assistant", "content": "
Title: ;
Story: ;
Question 1: ;
Answer 1: ;
Question 2: ;
Answer 2: ;
Question 3: ;
Answer 3: ;
Question 4: ;
Answer 4: ;
Question 5: ;
Answer 5: ;
Recap Question: ;
Recap Answer: ;
END"}

\subsection{Adjustment Prompt}
Easier: Please generate an easier version of this story, which could be by using more straightforward sentences, more basic vocabulary, simpler language, removing plot twists, clearer and predictable scenarios, or an easy-to-follow plot.
After the story, please generate five questions that can be used to communicate with the child and their answers
to review the emotional social scenarios and solutions in the story, with the recap question related to the child’s
own emotions in the story. The results are output according to the following format:"role": "user", "content":
"Emotion: Character: Item: Story Difficulty: ", "role": "assistant", "content": " Title: ; Story: ; Question 1: ; Answer
1: ; Question 2: ; Answer 2: ; Question 3: ; Answer 3: ; Question 4: ; Answer 4: ; Question 5: ; Answer 5: ; Recap
Question: ; Recap Answer: ; END"
More difficult: 
Please generate a more difficult version of this story, which could be by richer vocabulary, more advanced language, adding plot twists, more intricate scenarios, sophisticated emotional management processes, or varied social interactions.
After the story, please generate five questions that can be used to communicate with the child and their answers
to review the emotional social scenarios and solutions in the story, with the recap question related to the child’s
own emotions in the story. The results are output according to the following format:"role": "user", "content":
"Emotion: Character: Item: Story Difficulty: ", "role": "assistant", "content": " Title: ; Story: ; Question 1: ; Answer
1: ; Question 2: ; Answer 2: ; Question 3: ; Answer 3: ; Question 4: ; Answer 4: ; Question 5: ; Answer 5: ; Recap
Question: ; Recap Answer: ; END"

\subsection{Additional Consideration Prompt}
When generating stories, please pay attention to the following:
i) Use language that is easy to understand for children. Avoid abstract terms, and keep each sentence short in 20 words.
ii) Ensure the story has logical coherence with clear cause-and-effect relationships.
iii) Make the story realistic and close to the children's daily lives, avoiding abstract concepts.
iv) Try positive examples in the story to guide children in a positive direction.
v) The factors that trigger emotions should be accurate, simple, and direct.
vi) The story should be at least 50 words long to provide sufficient background for children to understand the emotions in the story.
vii) The sequence of questions should align with the development of the story.
viii) Emphasize the perspective of the main character when describing emotions.
ix) Appropriately include descriptions of emotional expressions and feelings to help children understand emotions.
x) To create stories that not only cognitively understand emotions, consider the following key points:
1) Emotional Resonance: Choose story themes that resonate with children, such as friendships and family relationships, to help them emotionally invest in the story.
2) Emotional Description: Describe the characters' emotional experiences in detail, including not only their actions but also their inner feelings, thoughts, and physical reactions.
3) Interactive Elements: Incorporate interactive aspects, such as asking children to imagine themselves as characters in the story and asking how they would feel and react.
4) Multisensory Experience: Use images, music, or other sensory elements to enhance the emotional expression in the story.
5) Emotional Guidance: Include questions that guide children to recognize and express their own emotions, helping them better understand and experience the emotions in the story.
6) Life Connection: Link the story plot to the children's real-life experiences, encouraging them to find similar emotional experiences in their daily lives.